\documentclass[aps,twocolumn,pra,nofootinbib,longbibliography]{revtex4-1}
\usepackage{amsmath,amssymb,bm}
\usepackage{graphicx}
\usepackage{epstopdf}
\usepackage{latexsym}
\usepackage{subfigure}
\usepackage[usenames,dvipsnames]{color}
\usepackage{hyperref}
\usepackage{natbib}
\usepackage{moreverb}
\begin{document}

\newcommand{\ns}{\mathcal{N}_{\mathrm{s}}}
\newcommand{\sinc}{\mathrm{sinc}}
\newcommand{\nb}{\mathcal{N}_{\mathrm{b}}}
\newcommand{\warn}[1]{{\color{red}\textbf{* #1 *}}}

\newcommand{\eqplan}[1]{{\color{blue}\textbf{equations:{ #1 }}}}

\newcommand{\figplan}[1]{{\color{blue}\textbf{figures: { #1 }}}}

\newcommand{\tableplan}[1]{{\color{blue}\textbf{tables: { #1 }}}}

\newcommand{\warntoedit}[1]{{\color{blue}\textbf{EDIT: #1 }}}

\newcommand{\warncite}[1]{{\color{green}\textbf{cite #1}}}

\newcommand{\mytitle}{Boson-mediated quantum spin simulators in transverse fields:XY model and spin-boson entanglement}

\title{\mytitle}
\date{\today}
\author{Michael L. Wall}
\affiliation{JILA, NIST and University of Colorado, 440 UCB, Boulder, CO 80309, USA}
\affiliation{Center for Theory of Quantum Matter, University of Colorado, Boulder, Colorado 80309, USA}
\author{ Arghavan Safavi-Naini}
\affiliation{JILA, NIST and University of Colorado, 440 UCB, Boulder, CO 80309, USA}
\affiliation{Center for Theory of Quantum Matter, University of Colorado, Boulder, Colorado 80309, USA}
\author{Ana Maria Rey}
\affiliation{JILA, NIST and University of Colorado, 440 UCB, Boulder, CO 80309, USA}
\affiliation{Center for Theory of Quantum Matter, University of Colorado, Boulder, Colorado 80309, USA}

\begin{abstract}

The coupling of spins to long-wavelength bosonic modes is a prominent means to engineer long-range spin-spin interactions, and has been realized in a variety of platforms, such as atoms in optical cavities and trapped ions.  To date, much of the experimental focus has been on the realization of long-range Ising models, but generalizations to other spin models are highly desirable.  In this work, we explore a previously unappreciated connection between the realization of an XY model by off-resonant driving of single sideband of boson excitation (i.e.~a single-beam M{\o}lmer-S{\o}rensen scheme) and a boson-mediated Ising simulator in the presence of a transverse field.  In particular, we show that these two schemes have the same effective Hamiltonian in suitably defined rotating frames, and analyze the emergent effective XY spin model through truncated Magnus series and numerical simulations.  In addition to XY spin-spin interactions that can be non-perturbatively renormalized from the naive Ising spin-spin coupling constants, we find an effective transverse field that is dependent on the thermal energy of the bosons, as well as other spin-boson couplings that cause spin-boson entanglement not to vanish at any time.  In the case of a boson-mediated Ising simulator with transverse field, we discuss the crossover from transverse-field Ising-like to XY-like spin behavior as a function of field strength.

\end{abstract}

\maketitle

\section{Introduction}

Externally driving a system with electromagnetic fields has become a ubiquitous means for engineering properties of synthetic many-body quantum systems that are difficult to obtain statically, with examples including the generation of synthetic gauge fields~\cite{PhysRevLett.109.145301,PhysRevLett.111.185301,PhysRevLett.111.185302,PhysRevX.4.031027,PhysRevLett.107.150501}, frustrated magnetism~\cite{struck2011quantum,struck2013engineering}, topological phases~\cite{PhysRevLett.94.086803,jotzu2014experimental,aidelsburger2015measuring,goldman2016topological}, and Floquet topological insulators~\cite{PhysRevB.82.235114,wang2013observation,rechtsman2013photonic}.  Often, the driving is used to generate a potential with desired single-particle properties, e.g., a topologically non-trivial band structure~\cite{PhysRevA.90.051601}, with the ultimate aim of combining this single-particle potential with interactions to study novel interacting quantum phases.  Theoretical analysis of such driven quantum systems is frequently based on a high-frequency expansion of the driven dynamics~\cite{PhysRevLett.116.125301,PhysRevX.4.031027,doi:10.1080/00018732.2015.1055918,PhysRevA.91.033632}.  When the drive is simultaneously applied with other non-commuting terms, e.g., interactions or other static fields, the applicability of the high-frequency limit and any conclusions drawn from it must be carefully re-examined~\cite{PhysRevA.90.043613,PhysRevB.91.245135}.

\begin{figure*}
\centering
\includegraphics[width=1.5\columnwidth]{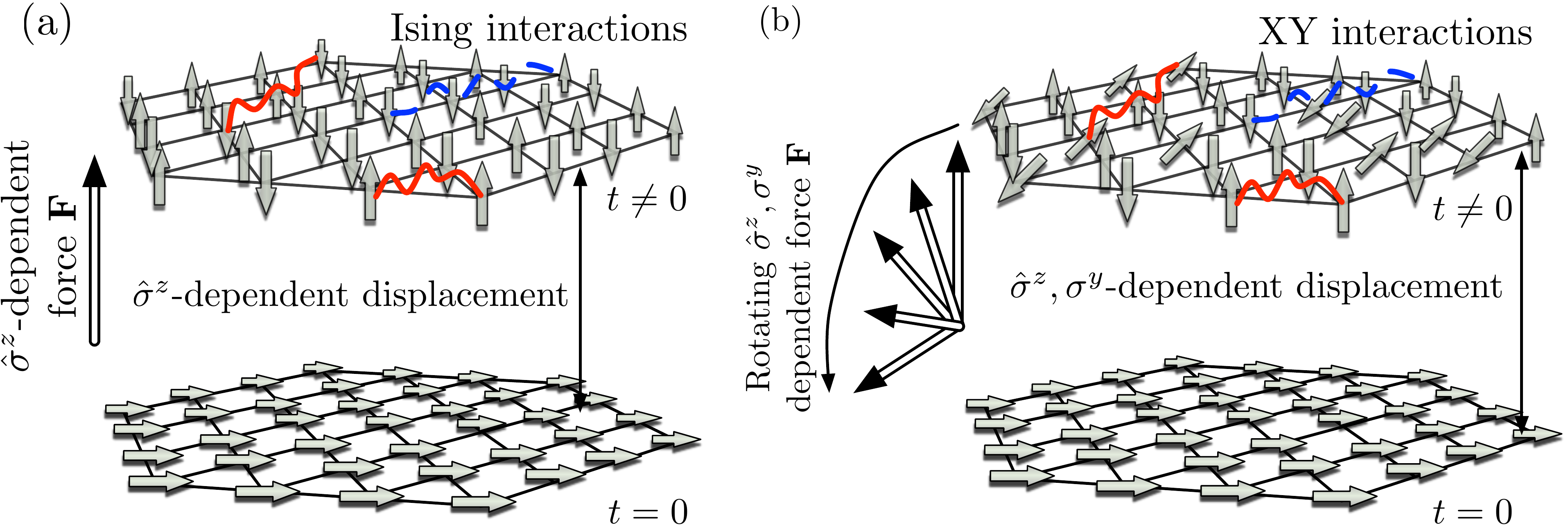}
\caption{(Color online)  \emph{Driven quantum Ising and XY spin simulators from static and rotating spin-dependent forces}  (a) Coupling of spins to boson modes via a spin-dependent force leads to spin-boson entanglement in the form of spin-dependent boson displacement (one possible spin configuration/displacement shown for clarity) and Ising spin-spin interactions that are positive (red solid) for aligned spins and negative (blue dashed) for anti-aligned spins.  (b) A rotating spin-dependent force, as occurs for an Ising simulator in the presence of a transverse field, leads to XY spin-spin interactions and qualitative changes to the spin-boson entanglement.}
\label{fig:Schematic}
\end{figure*}

One particularly successful example of realizing tailored many-body systems in this way is quantum spin systems coupled to long-wavelength boson modes with a spin-dependent drive.  Such coupled spin-boson systems can be realized in many platforms, such as cavity QED with atoms in optical cavities~\cite{baumann2010dicke,mottl2012roton,landig2016quantum} or superconducting qubit-based artificial atoms in microwave cavities~\cite{niemczyk2010circuit}, or in trapped ions, in which ion spins are coupled to phonon modes of the equilibrium crystal structure~\cite{PhysRevA.62.022311,PhysRevLett.82.1971,PhysRevLett.92.207901,PhysRevLett.103.120502,friedenauer2008simulating,britton2012engineered,PhysRevA.78.010101}.  In the absence of any additional fields, the high frequency expansion of a spin-dependent drive ($\propto \hat{\sigma}^z$) leads to an exact, terminating series featuring boson-mediated long-range Ising spin-spin interactions (Fig.~\ref{fig:Schematic}(a)).  This has led to many spectacular successes in quantum simulation of long-range interacting spin models~\cite{friedenauer2008simulating,kim2010quantum, PhysRevB.82.060412,britton2012engineered,richerme2014non,jurcevic2014quasiparticle,bohnet2015quantum}, which, in spite of the classical nature of the eigenstates of the Ising model, reveal non-classical features such as spin squeezing~\cite{PhysRevA.47.5138} in out-of-equilibrium dynamics.  In this same setting, it can also be shown that residual spin-boson entanglement, which degrades the fidelity of quantum simulation, can be made small or stroboscopically vanishing~\cite{Dylewsky,bohnet2015quantum}.

From the quantum simulation viewpoint, there is great experimental impetus to go beyond Ising spin-spin interactions and realize a richer class of models.  For example, many works have focused on adding an effective transverse field to the Ising simulator, ideally realizing a long-range transverse-field Ising model (TFIM) for which analytic solutions are unavailable and numerical simulations are challenging~\cite{islam2011onset,islam2013emergence,PhysRevLett.111.100506,PhysRevA.88.012334,richerme2014non,jurcevic2014quasiparticle,senko2014coherent}.  More complex spin-spin interactions could in principle be realized by coupling beams with different spin coupling character (e.g., $\propto \hat{\sigma}^x$ and $\propto \hat{\sigma}^y$) to distinct boson modes~\cite{PhysRevLett.92.207901}; however, experimentally this is challenging and has yet to be achieved.  Other proposals for more complex spin-spin interactions have arisen using only a single branch of boson modes, for example realizing an XY model from an assumed TFIM description with a static field~\cite{richerme2014non,jurcevic2014quasiparticle} or more complex models assuming drives with additional phase control and time-modulation~\cite{grass2016dual,bermudez2016long}.  However, in all of these more complex scenarios we lose the benefit of an exact high-frequency expansion, and the validity of mappings between the driven spin-boson system and effective spin-only descriptions depend on the particular parameters and timescales considered.

In this work, we explore in detail a particular means to generate more complex spin-spin interactions, namely an XY spin model, through the use of a spin-dependent force whose direction rotates in time (Fig.~\ref{fig:Schematic}(b)).  In particular, we show that the same XY description arises from two seemingly distinct physical realizations: (i) by applying a spin-dependent force which by itself stroboscopically generates an Ising model and  superimposing  an  additional transverse field and (ii) by driving of off-resonant spin flips near a single excitation sideband (i.e.~``half" of a M{\o}lmer-S{\o}rensen drive~\cite{PhysRevA.62.022311,PhysRevLett.82.1971}).  We show below that for both of these situations there exists an appropriate rotating frame where the high frequency expansion of the Hamiltonian yields an XY model.  Beyond effective spin-spin dynamics, we examine the coupling between spins and bosons, and show that terms exist in the high frequency expansion which couple spins and bosons and do not vanish at any time. 

Our analysis of the driven spin-boson system uses complementary analytical and numerical techniques.  On the analytical side, we derive effective spin models using truncated Magnus series which corresponds to a high-frequency limit in the detuning relative to the drive strength, but can be non-perturbative in the transverse field strength.  In addition, we look at the limit of weak transverse field using perturbative techniques.  We benchmark our analytical predictions against unbiased numerical simulations.

This work is organized as follows. In Sec.~\ref{sec:Models} we discuss the two microscopic models of spins coupled to bosons considered in this work: (i) quantum Ising spin simulators realized with spins coupled to bosons via spin-dependent forces in a superimposed transverse field and (ii) a single-frequency drive of spin excitations through off-resonant boson excitation.  In particular, after discussing different physical realizations of these two models, we show that they have the same effective Hamiltonian of a spin-dependent force whose spin direction rotates in time, in suitably defined rotating frames.  Sec.~\ref{sec:effXYOverview} presents an approximate expression for the time-ordered propagator of this effective Hamiltonian, based on a truncated Magnus series, and discusses its physical consequences, including an effective description in terms of a long-range XY spin model.  The validity of this approximate description is explored in Sec.~\ref{eq:XYRegimes} through numerical simulations with many spins and a single boson mode.  Sec.~\ref{sec:Fail} studies the parameter regimes where the XY model description fails in more detail, discusses alternate models for these regimes, and discusses how our XY model description smoothly connects with these descriptions.  Finally, Sec.~\ref{sec:Concl} concludes.  Detailed calculations of our truncated Magnus series appear as appendices.

\section{Driven spin-boson systems} 
\label{sec:Models}

In what follows, we consider two particular realizations of long-range spin models which arise by driving a collection of boson modes in a spin-dependent fashion.  In spite of their different physical realizations, we show that they have the same effective Hamiltonian in suitably defined rotating frames, and so display the same boson-mediated spin interactions in the high-frequency driving limit.

\subsection{Boson-mediated Ising simulator in a transverse field}
The first model we consider is a system of spin 1/2 particles driven by external fields that generate a spin dependent force in the presence of an additional transverse field.  For concreteness we will focus on realizations of this Hamiltonian in trapped ions, and point out related descriptions in other systems momentarily.

Spin-dependent forces are engineered in trapped ion experiments through two different mechanisms, which we now briefly review for future reference.  In the first, the spin-dependent force arises from an AC Stark shift due to a running-wave optical lattice formed by the interference of two laser beams with Raman beatnote frequency $\omega_R$~\cite{leibfried2003experimental,britton2012engineered}.  The Hamiltonian describing a system of spins and phonons coupled in this fashion at lowest order in the Lamb-Dicke expansion is ($\hbar=1$ throughout) 
\begin{align}
\label{eq:H} \hat{H}\left(t\right)&=\hat{H}_{\omega}+\hat{H}_{\mathrm{SB}}+\hat{H}_{\mathrm{qubit}}\, ,\\
\hat{H}_{\omega}&=\sum_{\mu}\omega_{\mu}\hat{n}_{\mu}\, ,\\
\hat{H}_{\mathrm{qubit}}&=\frac{\omega_{\mathrm{eg}}}{2}\sum_i \hat{\sigma}^z_i\, ,\\
\hat{H}_{\mathrm{SB}}&=-\sum_{\mu j} g_{\mu j} \cos\left(\omega_{R} t\right) \left(\hat{a}_{\mu}+\hat{a}^{\dagger}_{\mu}\right)\hat{\sigma}^z_j\, .
\end{align} 
The operators $\hat{\sigma}^{x,y,z}_i$ are Pauli operators acting on the spin-1/2 or qubit degree of freedom, $\hat{a}_{\mu}$ is an anihilation operator for boson mode $\mu$, and $\hat{n}_{\mu}=\hat{a}^{\dagger}_{\mu}\hat{a}_{\mu}$.  In the trapped ion realization~\cite{PhysRevLett.92.207901,britton2012engineered}, $\{\omega_{\mu}\}$ are the phonon mode frequencies, $\omega_{\mathrm{eg}}$ is the ``bare" ion qubit frequency, and $g_{\mu j}=Fb_{j\mu}\sqrt{\hbar/(2M\omega_{\mu})}$ codifies the force on the $j^{\mathrm{th}}$ ion due to driving the $\mu^{\mathrm{th}}$ phonon, with $F$ the magnitude of the spin-dependent force, $\mathbf{b}_{\mu}$ the normalized mode amplitude of phonon $\mu$, and $M$ the ion mass.  The beams are arranged in a geometry such that the net momentum transfer is nonzero only along a particular direction with respect to the ion crystal-exciting the phonon modes described by $\hat{a}_{\mu}$-and the polarization of the beams is chosen such that the spin-independent differential Stark shift vanishes and only a spin-dependent shift remains.  

Since the single-spin Hamiltonian commutes with the spin-boson coupling, we can transform to a frame rotating with $\hat{H}_{\mathrm{qubit}}$ without changing the form of the spin-boson coupling Hamiltonian.  In addition, it is useful to transform to the interaction picture rotating with $\hat{H}_{\omega}$ and perform a rotating wave approximation.  This (optical) rotating wave approximation is not essential, but considerably simplifies expressions, and is an excellent approximation in the parameter regimes explored herein.  Finally, we additionally consider the addition of an effective transverse field, which is realized through direct coupling of the two states realizing the spin-1/2 degree of freedom (e.g.~microwave coupling of ion qubits).  The complete Hamiltonian of the spin-dependent force with the transverse field in this frame is then
\begin{align}
\label{eq:HI} \hat{H}_I\left(t\right)&=\hat{H}_{\mathrm{SB};I}\left(t\right)+\hat{H}_B\, ,\\
\label{eq:HISB}\hat{H}_{\mathrm{SB};I}\left(t\right)&=-\frac{1}{2}\sum_{\mu j} g_{\mu j} \left(\hat{a}^{\dagger}_{\mu}e^{-i\delta_{\mu} t}+\hat{a}_{\mu}e^{i\delta_{\mu} t}\right)\hat{\sigma}^z_j\, ,\\
\hat{H}_B&=-\frac{B}{2}\sum_j\hat{\sigma}^x_j\, ,
\end{align}
where $\delta_{\mu}=\omega_R-\omega_{\mu}$ is the detuning of the drive from mode $\mu$.

The second common realization of a spin-dependent force uses the M{\o}lmer-S{\o}rensen (MS) scheme~\cite{PhysRevLett.82.1971,PhysRevA.62.022311}.  Here, the Hamiltonian takes the same form of Eq.~\eqref{eq:H}, with the spin-phonon coupling given by a pair of Raman beams described by the Hamiltonian
\begin{align}
\label{eq:bareMS}\hat{H}_{\mathrm{SB}}&=\sum_{q}\sum_j\frac{\Omega_{q}}{2}\left(\hat{\sigma}^+_je^{i\left[\sum_{\mu}\eta_{\mu j;q}\left(\hat{a}_{\mu}+\hat{a}^{\dagger}_{\mu}\right)+\left(\omega_{\mathrm{eg}}-\omega_q\right)t\right]}+\mathrm{H.c.}\right)\, .
\end{align}
In this expression, $q=1,2$ indexes the Raman beam pairs, $\Omega_{q}$ is the resonant Rabi frequency of beam $q$, $\eta_{\mu j;q}$ is the Lamb-Dicke parameter $k_q b_{j\mu}\sqrt{\hbar/(2M\omega_{\mu})}$ with $k_q$ the wavevector of the Raman beam pair $q$, and $\omega_q$ is the frequency of beam pair $q$, defined with respect to the bare qubit frequency.  In particular, in the MS scheme the beams have the same Rabi frequencies $\Omega_{1}=\Omega_{2}=\Omega$ and equal but opposite detuning, $\omega_1=\omega_{\mathrm{bsb}}=\omega_{\mathrm{eg}}+\omega_{\mu}+\delta_{\mu}$, $\omega_2=\omega_{\mathrm{rsb}}=\omega_{\mathrm{eg}}-\omega_{\mu}-\delta_{\mu}$.  Hence, beam 1 corresponds to a blue sideband excitation and beam 2 to a red sideband excitation.  Transferring to the rotating frames of $\hat{H}_{\omega}$ and $\hat{H}_{\mathrm{qubit}}$, expanding the exponentials to lowest order in the Lamb-Dicke parameters $\eta_{\mu j;1}\approx \eta_{\mu j;2}\equiv \eta_{\mu j}$, and performing a rotating wave approximation where we neglect terms with frequencies larger than the detunings $\delta_{\mu}$, we find
\begin{align}
\hat{H}_{\mathrm{SB};I}&=-\sum_j\sum_{\mu} \frac{\Omega\eta_{\mu j}}{2}\left(e^{-i\delta_{\mu} t}\hat{a}^{\dagger}_{\mu}+e^{i\delta_{\mu} t}\hat{a}_{\mu}\right)\hat{\sigma}^y_j\, .
\end{align}
Hence, the MS scheme reproduces exactly the interaction picture Hamiltonian of Eq.~\eqref{eq:HISB} following a trivial rotation of the spin basis which maps $\hat{\sigma}^y_j\to\hat{\sigma}^z_j$ and the identification of $g_{\mu j}\equiv \Omega \eta_{\mu j}$.  More generally, after performing a spin rotation, Eqs.~\eqref{eq:HI}-\eqref{eq:HISB} describe Rabi-type atom-photon interactions, with $g_{\mu j}$ the atom-photon coupling strength and $B$ the frequency of the atomic transition.  Hence, our analysis of this model also applies to artificial cavity QED systems in which cavity dissipation may be neglected.

We refer to the Hamiltonian Eq.~\eqref{eq:HI} as a driven spin-boson Ising simulator in a transverse field because the propagator when $\hat{H}_B\to 0$ is exactly~\cite{PhysRevLett.82.1971,Dylewsky}
\begin{align}
\label{eq:IsingMagExact}\hat{U}_I\left(t\right)&=\hat{U}_{\mathrm{SB}}\left(t\right)\hat{U}_{\mathrm{SS}}\left(t\right)
\end{align}
with the spin-boson and spin-spin coupling propagators
\begin{align}
\hat{U}_{\mathrm{SB}}\left(t\right)&= \exp[\sum_{\mu,j} g_{\mu j}\left(\alpha_{\mu}\left(t\right)\hat{a}_{\mu}^{\dagger}-\mathrm{H.c.}\right)\hat{\sigma}^z_j]\,,\\
\hat{U}_{\mathrm{SS}}\left(t\right)&=\exp[-i \sum_{i,j}\tilde{J}_{ij}\left(t\right)\hat{\sigma}^z_i\hat{\sigma}^z_j]\, ,
\end{align}
and spin-boson and spin-spin coupling parameters
\begin{align}
\label{eq:alp} \alpha_{\mu}(t)&=(1-e^{-i\delta_{\mu} t})/(2\delta_{\mu})\, ,\\
\label{eq:Jtil} \tilde{J}_{jj'}\left(t\right)&=\sum_{\mu} g_{\mu j}g_{\mu j'}(\delta_{\mu} t-\sin(\delta_{\mu} t))/(4\delta_{\mu}^2)\, .
\end{align}
At times long compared to $1/\delta$ where $\delta\equiv \min_{\mu}\delta_{\mu}$, $\tilde{J}_{jj'}\left(t\right)$ can be approximated by its unbounded in time (secular) component as $\tilde{J}_{jj'}\left(t\right)\approx J_{jj'} t$, where $J_{jj'}=\sum_{\mu} g_{\mu j}g_{\mu j'}/(4\delta_{\mu})$.  In this approximation, $\hat{U}_{\mathrm{SS}}\left(t\right)$ is the propagator of a long-range Ising model $\hat{H}_{\mathrm{Ising}}=\sum_{i,j}J_{i,j}\hat{\sigma}^z_i\hat{\sigma}^z_j$.  The spin-spin couplings can be approximated by a power-law form $J_{i,j}\sim 1/|i-j|^{\zeta}$, with the exponent $\zeta\in\left[0,3\right)$ being tunable by the Raman beatnote frequency $\omega_R$~\cite{britton2012engineered,PhysRevA.92.043405}.  For a single mode with detuning $\delta_{\mu}$, the spins decouple from the bosons at the decoupling times $t_d=2\pi n/\delta_{\mu}$, $n$ an integer, where $\alpha_{\mu}(t_d)=0$.  In the case where many modes contribute to the dynamics, the various detunings $\{\delta_{\mu}\}$ are not generally commensurate, and the only means to approximately decouple the spins from the bosons is to have the amplitude of the spin-boson couplings $g_{\mu j}\alpha_{\mu}(t)$ parametrically small, i.e., $g_{\mu j}/\delta_{\mu}$ must be small~\cite{Dylewsky}.  

In many works~\cite{islam2011onset,islam2013emergence,PhysRevLett.111.100506,PhysRevA.88.012334,richerme2014non,jurcevic2014quasiparticle,senko2014coherent,PhysRevA.86.032329}, the small-$B$ limit of Eq.~\eqref{eq:HI} has been considered, and the physics is shown to be well-described by the transverse-field Ising model 
\begin{align}
\label{eq:TFIM}\hat{H}_{\mathrm{TFIM}}&=\sum_{j,j'}J_{j,j'}\hat{\sigma}^z_j\hat{\sigma}^z_{j'}-\frac{B}{2}\sum_{j}\hat{\sigma}^x_j\, ,
\end{align}
that is derived from treating $\hat{H}_{\mathrm{B}}$ perturbatively in the high-frequency expansion and ignoring the boson dynamics.   However, importantly, this high-frequency expansion no longer exactly truncates due to the non-commutativity of $\hat{H}_B$ and $\hat{H}_{\mathrm{SB};I}$.  This means that higher-order terms in the expansion can modify, even qualitatively, both the spin physics and the spin-boson coupling.  In order to derive a high-frequency expansion of the time-ordered dynamics of Eq.~\eqref{eq:HI} which is non-perturbative in the field strength $B$, it is convenient to transform to a frame which rotates with the transverse field Hamiltonian $\hat{H}_B$.  The Hamiltonian in this frame reads
\begin{align}
\nonumber \hat{\mathcal{H}}_I\left(t\right)&=-\sum_{\mu j}\frac{g_{\mu j}}{2}\left(\hat{a}_{\mu}e^{i\delta_{\mu} t}+\hat{a}_{\mu}^{\dagger}e^{-i\delta_{\mu} t}\right)\\
\label{eq:IPwithB}&\times \left(\cos\left(B t\right)\hat{\sigma}^z_j-\sin\left(B t\right)\hat{\sigma}^y_j\right)\, .
\end{align}
This is indeed a spin dependent force whose spin character rotates between the $z$ and $y$ directions at a rate $B$ (Fig.~\ref{fig:Schematic}(b)).  A discussion on the boson-mediated spin physics derived from Eq.~\eqref{eq:IPwithB} will be presented in Sec.~\ref{sec:effXYOverview}, and is \emph{not} a transverse-field Ising model for all parameter regimes.

\subsection{Single-beam M{\o}lmer-S{\o}rensen scheme}
The second model that we consider is a modified M{\o}lmer-S{\o}rensen scheme in which only \emph{one} of the pairs of Raman beams is present.  For 
concreteness, let us consider that only the blue sideband beams are present, in which case we have the Hamiltonian (see Eq.~\eqref{eq:bareMS})
\begin{align}
\nonumber  \hat{H}_{\mathrm{SB};I}&=
 \sum_j\frac{\Omega}{2}\left(\cos\left(\omega_{\mathrm{bsb}} t\right)\hat{\sigma}^x_j+\sin\left(\omega_{\mathrm{bsb}} t\right)\hat{\sigma}^y_j\right)\\
 \nonumber &-\sum_j\sum_{\mu} \frac{\Omega\eta_{\mu j}}{2}\left(\cos\left(\omega_{\mathrm{bsb}} t\right)\hat{\sigma}^y_j-\sin\left(\omega_{\mathrm{bsb}} t\right)\hat{\sigma}^x_j\right)\\
 \label{eq:OnebeamMS}&\times \left(\hat{a}^{\dagger}_{\mu}e^{i\omega_{\mu} t}+\hat{a}_{\mu}e^{-i\omega_{\mu} t}\right)\, ,
\end{align}
where, as before, $\omega_{\mathrm{bsb}}=\omega_{\mathrm{eg}}+\omega_{\mu}+\delta_{\mu}$ and we are in the frame rotating with $\hat{H}_{\omega}$ and $\hat{H}_{\mathrm{qubit}}$.  The first term in Eq.~\eqref{eq:OnebeamMS} is a spin rotation that is not coupled to the bosons, and gives rise to an AC Stark shift in perturbation theory.  In order to put Eq.~\eqref{eq:OnebeamMS} into the form of Eq.~\eqref{eq:IPwithB}, we first move to a frame which rotates with $({\omega_{\mathrm{bsb}}}/{2})\sum_{j}\hat{\sigma}^z_j$, where the effective Hamiltonian reads
\begin{align}
\nonumber \hat{H}_{\mathrm{SB};I'}&=-\frac{\omega_{\mathrm{bsb}}}{2}\sum_j \hat{\sigma}^z_j+\sum_j\frac{\Omega}{2}\hat{\sigma}^x_j\\
&+\sum_j\sum_{\mu} \frac{\Omega\eta_{\mu j}}{2}\hat{\sigma}^y_j\left(\hat{a}^{\dagger}_{\mu}e^{i\omega_{\mu} t}+\hat{a}_{\mu}e^{-i\omega_{\mu} t}\right)\, .
\end{align}
If we now move to a frame that rotates with $-({\omega_{\mathrm{bsb}}}/{2})\sum_j \hat{\sigma}^z_j+\sum_j({\Omega}/{2})\hat{\sigma}^x_j$ we find the effective Hamiltonian
\begin{align}
\nonumber &\hat{H}_{\mathrm{SB};I''}=\sum_j\sum_{\mu} \frac{\Omega\eta_{\mu j}}{2}\left(\hat{a}^{\dagger}_{\mu}e^{i\omega_{\mu} t}+\hat{a}_{\mu}e^{-i\omega_{\mu} t}\right)\\
\label{eq:MSfinaleff}&\times \left[\cos\left(\omega_{\mathrm{eff};j} t\right)\hat{\sigma}^y_j-\hat{\boldsymbol{\sigma}}\cdot\mathbf{n} \sin\left(\omega_{\mathrm{eff};j} t\right)\right]\, ,
\end{align}
where $\omega_{\mathrm{eff};j}=\sqrt{\omega_{\mathrm{bsb}}^2+\Omega^2}$ and the unit vector $\mathbf{n}=(\omega_{\mathrm{bsb}},0,-\Omega )/{\omega_{\mathrm{eff};j}}$.  

One may expect that the first term in Eq.~\eqref{eq:OnebeamMS} is irrelevant on the basis that it rotates fast compared to the second term, and so may be ignored.  Our rotating frame analysis above enables us to make this intuition more precise, as follows. The operator which defines the interaction picture in which the effective Hamiltonian Eq.~\eqref{eq:MSfinaleff} applies is
\begin{align}
\hat{U}_{I''}&=\prod_j \hat{U}_j\\
\label{eq:Ujs}\hat{U}_j&=e^{-i\frac{\omega_{\mathrm{bsb}}t}{2} \hat{\sigma}^z_j+i\frac{\Omega t}{2}\hat{\sigma}^x_j}e^{i\frac{\omega_{\mathrm{bsb}}t}{2}\hat{\sigma}^z_j}\, .
\end{align}
Clearly, as $\Omega\to 0$, $\hat{U}_j$ becomes the identity for all times.  In the basis of $\hat{\sigma}^z_j$, the diagonal elements of $\hat{U}_j$ consist of a terms rotating as $\exp(\pm i\Omega^2 t/\omega_{\mathrm{bsb}})$ which have order unity amplitudes up to $\mathcal{O}(\Omega^2/\omega_{\mathrm{bsb}}^2)$ corrections and terms rotating as $\exp(\pm it(2\omega_{\mathrm{bsb}}+\Omega^2 /\omega_{\mathrm{bsb}}))$ which have order $\mathcal{O}(\Omega^2/\omega_{\mathrm{bsb}}^2)$ amplitudes.  The off-diagonal components also contain terms rotating as $\exp(\pm i\Omega^2 t/\omega_{\mathrm{bsb}})$ and $\exp(\pm it(2\omega_{\mathrm{bsb}}+\Omega^2 /\omega_{\mathrm{bsb}}))$, all of which have order $\mathcal{O}(\Omega/\omega_{\mathrm{bsb}})$ amplitudes.  In this rotating frame, the effective Hamiltonian now has the same form as Eq.~\eqref{eq:IPwithB} provided we perform a spin rotation to take $\hat{\boldsymbol{\sigma}}\cdot\mathbf{n} \to \hat{\sigma}^y_j$ and $\hat{\sigma}^y_j\to \hat{\sigma}^x_j$ and identify $\delta_{\mu}\to-\omega_{\mu}$, $B\to \omega_{\mathrm{eff}}$, and $g_{\mu j}\to-\Omega\eta_{\mu j}$.  We note that the case of a single-beam MS scheme has been considered before in the context of spin models~\cite{PhysRevLett.90.133601} and has been experimentally utilized to engineer a spin-one XY model~\cite{PhysRevX.5.021026}.

\section{Effective XY model: overview} 
\label{sec:effXYOverview}
Our model of a rotating spin-dependent force Eq.~\eqref{eq:IPwithB}, while compact and superficially simple, does not immediately enable us to determine the dominant boson-mediated spin-spin physics.  In addition, as this Hamiltonian is explicitly time-dependent, our analysis of the spin-spin interactions must take into account the necessary time-ordering.  A systematic means of determining properly time-ordered, unitary approximations to the propagator of a time-dependent Hamiltonian is provided by the exponential of the Magnus series~\cite{Magnus_54,Blanes_Casas_09} $\hat{U}(t)=\exp(\hat{\mathcal{A}}(t))$, where $\hat{\mathcal{A}}\left(t\right)=\sum_{k=1}^{\infty}\hat{\mathcal{A}}_k\left(t\right)$ is a sum of integrals $\hat{\mathcal{A}}_k$ of $k$ nested commutators of the Hamiltonian with itself at different times (see Eq.~\eqref{eq:MagRecur} for an explicit expression).  The Magnus series does not always converge for all times, but is guaranteed to converge at short times and often provides an efficient and accurate means to construct effective Hamiltonians over experimental timescales in the high-driving-frequency limit~\cite{doi:10.1080/00018732.2015.1055918}.

If we truncate the Magnus series generated by Eq.~\eqref{eq:IPwithB} at second order, we find
\begin{align}
\nonumber\hat{\mathcal{A}}(t)&\approx\sum_{\mu} \sum_j g_{\mu j} \left[\hat{a}^{\dagger}_{\mu}({\textstyle \sum_{\nu=z,y}}\alpha^{\nu}_{\mu}(t)\hat{\sigma}^{\nu}_j)-\mathrm{H.c}\right]\\
\nonumber&+\sum_{\mu,\mu'}\sum_j g_{\mu,j}g_{\mu',j}\hat{\sigma}^x_j\Big[\hat{a}^{\dagger}_{\mu}\hat{a}^{\dagger}_{\mu'}\alpha^{++}_{\mu\mu';j}(t)\\
\nonumber&+\left(1-\delta_{\mu,\mu'}\right)\hat{a}^{\dagger}_{\mu}\hat{a}_{\mu'}\alpha^{+-}_{\mu\mu';j}(t)-\mathrm{H.c.}\Big]\\
\nonumber&-i\sum_{j\ne j'}\left[\tilde{J}_{j,j'}^{zz}(t)\hat{\sigma}^z_{j}\hat{\sigma}^z_{j'}+\tilde{J}^{yy}_{j,j'}(t)\hat{\sigma}^y_{j}\hat{\sigma}^y_{j'}+\tilde{J}^{yz}_{j,j'}(t)\hat{\sigma}^y_{j}\hat{\sigma}^z_{j'}\right]\\
\label{eq:NPMagnus} &-i\sum_{\mu}\sum_jg_{\mu j}^2{B}_{\mathrm{eff};\mu}(t)\hat{\sigma}^x_j\left(2\hat{n}_{\mu}+1\right)\, ,
\end{align}
as is detailed in Appendix~\ref{app:Magnus}, where expressions for all coefficients may be found.  The identification of the model Eq.~\eqref{eq:NPMagnus} is one of our key results.  In what follows we discuss each of the terms in the model, as well as the qualitative behavior of its parameters with time, drive strength, transverse field strength, and driving frequency.  

While we use the notation of the boson-mediated Ising simulator in a transverse field in what follows, we remind the reader that the same model applies for the single-beam MS scheme following the mapping of parameters discussed following Eq.~\eqref{eq:Ujs}.  We also note that this Magnus expansion applies in rotating frames which are different for our two physical realizations.  Namely, the Schr\"{o}dinger picture evolution for the spin-boson Ising simulator in a transverse field is
\begin{align}
\label{eq:ModIMag}|\psi(t)\rangle&=\exp[i(\frac{B}{2}\sum_j \sigma^x_j-\sum_{\mu} \omega_{\mu}\hat{n}_{\mu} )t]\exp(\hat{\mathcal{A}}(t))|\psi(0)\rangle\, ,
\end{align}
and the evolution for the single-beam MS gate is
\begin{align}
\label{eq:ModIIMag}|\psi(t)\rangle&=\exp[-i\frac{\omega_{\mathrm{bsb}}t}{2}\sum_j \hat{\sigma}^z_j+i\sum_j\frac{\Omega t}{2}\hat{\sigma}^x_j]\\
\nonumber &\times\exp[i(\frac{\omega_{\mathrm{bsb}}}{2}\sum_j \sigma^z_j-\sum_{\mu} \omega_{\mu}\hat{n}_{\mu} )t]\exp(\hat{\mathcal{A}}(t))|\psi(0)\rangle\, .
\end{align}
As noted earlier, the spin rotation incurred by Eq.~\eqref{eq:ModIIMag} occurs with small amplitude $\Omega/\omega_{\mathrm{eff}}$ and at frequencies well separated from the frequencies of the dynamical evolution in $\hat{\mathcal{A}}(t)$.  In contrast, the spin rotation in Eq.~\eqref{eq:ModIMag} occurs at frequencies that are comparable to the dynamics of $\hat{\mathcal{A}}(t)$ and has unity amplitude.  Hence, generally speaking, we expect that the convergence criteria of the Magnus series to the exact solution for the first case Eq.~\eqref{eq:ModIMag} will be stricter than for Eq.~\eqref{eq:ModIIMag}.

 The first-order terms, given by the first line of Eq.~\eqref{eq:NPMagnus}, generate spin-boson entanglement via spin-dependent displacements whose spin direction rotates in time.  Both $\alpha^z_{\mu}(t)$ and $\alpha^y_{\mu}(t)$ are bounded and vanish stroboscopically at integer multiples of the modified decoupling time $\tilde{t}_d=2\pi/(p \delta_{\mu} )$ when $p \delta_{\mu}=Bq$ for coprime integers $p$ and $q$.  As $B\to 0$, $\alpha^y_{\mu}(t)\to0$ and $\alpha^z_{\mu}(t)\to \alpha_{\mu}(t)$ so that the exact, finite Magnus series for the boson-mediated Ising model Eq.~\eqref{eq:IsingMagExact} is reproduced, as expected.  For finite $B$, the maximum value of both $\alpha^z_{\mu}(t)$ and $\alpha^y_{\mu}(t)$ scales as $\max(\delta_{\mu},B)/(\delta_{\mu}^2-B^2)$.  The Magnus series also contains terms which couple pairs of bosons to spins; these are the second and third lines of Eq.~\eqref{eq:NPMagnus}.  These terms are also bounded and vanish at integer multiples of $\tilde{t}_d$.  As an example of their magnitude, the maximum value of $\alpha^{++}_{\mu,\mu}(t)$ scales as $\max(\delta_{\mu},B)/(\delta_{\mu}(\delta_{\mu}^2-B^2))$.
  
The fourth line of Eq.~\eqref{eq:NPMagnus} contains spin-spin interactions along the $zz$, $yy$, and $yz$ interactions.  The couplings $\tilde{J}^{yz}_{jj'}(t)$ are bounded, and vanish at multiples of $\tilde{t}_d$.  The $zz$ and $yy$ couplings are
\begin{align}
\label{eq:Jzz}\tilde{J}^{zz}_{j,j'}(t)&=\sum_{\mu} \frac{g_{\mu,j}g_{\mu,j'}\delta_{\mu}}{8}\frac{t+ \sin(2Bt)/2B}{\delta_{\mu}^2-B^2}+\mbox{b.t. }\, ,\\
\label{eq:Jyy}\tilde{J}^{yy}_{j,j'}(t)&=\sum_{\mu}\frac{g_{\mu j}g_{\mu j'}\delta_{\mu} }{8}\frac{t- \sin(2Bt)/2B}{\delta_{\mu}^2-B^2}+\mbox{b.t. }\, ,
\end{align} 
where $\mbox{b.t.}$ denotes bounded terms which oscillate at frequencies $\delta_{\mu}$ and $B$ and vanish at $\tilde{t}_d$.  For times $t\ll 1/B$, $\tilde{J}^{z,z}_{j,j'}(t)\approx \tilde{J}_{j,j'}(t)$ and $\tilde{J}^{y,y}_{j,j'}(t)\approx 0$, and (transforming back out of the rotating frame) we recover the TFIM Eq.~\eqref{eq:TFIM} at leading order in $Bt$.  However, at times $t\gtrsim B$, the secular terms in $\tilde{J}^{z,z}_{j,j'}(t)$ and $\tilde{J}^{y,y}_{j,j'}(t)$ dominate and are of equal strength, leading to a description directly in terms of a long-range XY spin model in the spin directions perpendicular to the transverse field.

 In addition to a modification of the operator character of the spin-spin couplings, we also find a non-perturbative renormalization of their strength.  As $B$ is increased relative to $\delta_{\mu}$, the spin-spin couplings change sign and their scaling changes from $\sim1/\delta_{\mu}$ to $\sim\delta_{\mu}/B^2$.  While it appears that the couplings diverge at the resonant point $B=\delta_{\mu}$, in fact the bounded terms regularize this divergence and lead to finite spin-spin couplings (further discussion of the resonant point is given in Sec.~\ref{sec:resonant}).  The final line of Eq.~\eqref{eq:NPMagnus} acts as an effective transverse field in the rotating frame.  This field is proportional to the thermal energy of the boson modes, and is generally spatially inhomogeneous--even for a spatially uniform external field $B$--by virtue of the coupling amplitudes $g_{\mu j}$.  The effective field strength is
\begin{align}
\nonumber  B_{\mathrm{eff};\mu}&=\frac{B t}{4\left(\delta_{\mu}^2-B^2\right)}+\frac{\left(\delta_{\mu}^2+B^2\right)\cos\left(\delta_{\mu} t\right)\sin\left(B t\right)}{4\left(\delta_{\mu}^2-B^2\right)^2}\\
 &-\frac{2B\delta_{\mu} \cos\left(B t\right)\sin\left(\delta_{\mu} t\right)}{4\left(\delta_{\mu}^2-B^2\right)^2}\, ,
\end{align} 
and hence contains a secular term of magnitude $Bt/(4(\delta_{\mu}^2-B^2))$.  It is interesting that in both the small $B/\delta_{\mu}$ and large $B/\delta_{\mu}$ limits $B_{\mathrm{eff};\mu}$ vanishes; in the first limit it vanishes as $\sim B$ and in the latter as $\sim 1/B$.

To second order, the Magnus operator $\hat{\mathcal{A}}(t)$ contains exact decoupling points $\tilde{t}_d$ for a single boson mode in a Fock state, analogous to the decoupling points $t_d$ of the pure driving Hamiltonian Eq.~\eqref{eq:HISB}.  However, at third order we find spin-boson coupling terms
\begin{align}
 \label{eq:thirdorderSP}&\sum_{j\mu} g_{\mu j}^3\Big[ \left\{\hat{a}_{\mu},\left(\hat{a}_{\mu}^{\dagger}\right)^2\right\}({\textstyle \sum_{\nu=z,y}}\alpha^{\nu\left(2,1\right)}_{\mu}(t)\hat{\sigma}^{\nu}_j)-\mathrm{H.c.}\Big]\, ,
\end{align}
in which $\alpha^{y\left(2,1\right)}_{\mu}$ and $\alpha^{z\left(2,1\right)}_{\mu}$ contain secular components which scale as $B\delta_{\mu}t/(2(\delta_{\mu}^2-4B^2)^2)$ and $B^2 t/(4(\delta_{\mu}^2-B^2)^2)$, respectively (Appendix~\ref{app:MagnusTO}).  These terms lead to non-vanishing spin-boson entanglement at times $t\gtrsim 1/B$, even for Fock states.  For certain limits, e.g. $\delta_{\mu} \gg g_{\mu,j},B$ or $B \gg g_{\mu,j},\delta_{\mu}$, the slope of the third-order secular terms can be made parametrically small to assuage the buildup of spin-boson entanglement.  In practice, due to the rapid decrease of the spin-spin coupling constants with $B$ in the limit $B \gg g_{\mu,j},\delta_{\mu}$, only the limit $\delta_{\mu} \gg g_{\mu,j},B$ produces negligible spin-boson entanglement and non-negligible spin-spin interactions simultaneously in this model.  We mention the term Eq.~\eqref{eq:thirdorderSP} only to give an example of a term in the Magnus series which leads to non-vanishing spin-boson entanglement at the decoupling points $\tilde{t}_d$; other terms also appear at third order, such as spin-spin-boson couplings (see Appendix~\ref{app:MagnusTO}).

Following the same analysis that converts the evolution under the spin-dependent force, Eq.~\eqref{eq:IsingMagExact}, into evolution under a long-range Ising model by keeping only secular terms we arrive at the effective spin model corresponding to the Magnus series of Eq.~\eqref{eq:NPMagnus},
\begin{align}
\label{eq:XYeff}\hat{H}_{\mathrm{XY}}&=\sum_{j\ne j'}J_{j,j'}^{\mathrm{XY}}\left[\hat{\sigma}^z_{j}\hat{\sigma}^z_{j'}+\hat{\sigma}^y_{j}\hat{\sigma}^y_{j'}\right]+\sum_j \mathcal{B}_j(n)\hat{\sigma}^x_j\, .
\end{align}
Here, $J_{j,j'}^{\mathrm{XY}}=\sum_{\mu} {g_{{\mu},j}g_{{\mu},j'}\delta_{{\mu}}}/{8(\delta_{{\mu}}^2-B^2)}$ and $\mathcal{B}_j(n)=\left(2n+1\right)\sum_{\mu}g_{{\mu} j}^2B/(4(\delta_{{\mu}}^2-B^2))$.  Here, $n$ is a $c$-number parameter which approximates the boson operator $\hat{n}$ (e.g., for the dynamics of an initial phonon Fock state one can set $n$ to the initial phonon number). In the next section, we quantitatively determine the accuracy of this effective description using numerical simulations.

\section{Effective XY model: Regimes and Validity} 
\label{eq:XYRegimes}
The above Magnus series Eq.~\eqref{eq:NPMagnus} is an expansion whose convergence properties are difficult to ascertain analytically.  Hence, in this section, we test this model, the idealized XY spin model Eq.~\eqref{eq:XYeff} derived from it, and the perturbative TFIM description Eq.~\eqref{eq:TFIM}, against unbiased numerical simulations.  For simplicity we present results for the single-mode case, and take the mode amplitude to be uniform: $g_{\mu}=g/\sqrt{N}$ with $N$ the number of particles.  In this scenario, all dynamics is restricted to occur in the tensor product of the completely symmetric Dicke spin manifold and the boson Hilbert space, with total dimension $(N+1)\times (N_{\mathrm{max}}+1)$, where $N_{\mathrm{max}}$ is the maximum boson occupation, here taken to be 50.  In addition to being theoretically convenient due to its small Hilbert space size, this scenario is also relevant for trapped ion experiments, corresponding to the case in which the drive frequency is close to the center of mass (COM) mode~\cite{britton2012engineered,bohnet2015quantum}.  The COM mode is the mode with the highest frequency, and can be well spectroscopically resolved.  We have also confirmed that the same qualitative behavior occurs in situations with many modes using a recently developed framework for generic driven spin-boson models~\cite{SBMPS} based on matrix product states (MPSs)~\cite{Schollwoeck}.  

\begin{figure}[t]
\centering
\includegraphics[width=0.9\columnwidth]{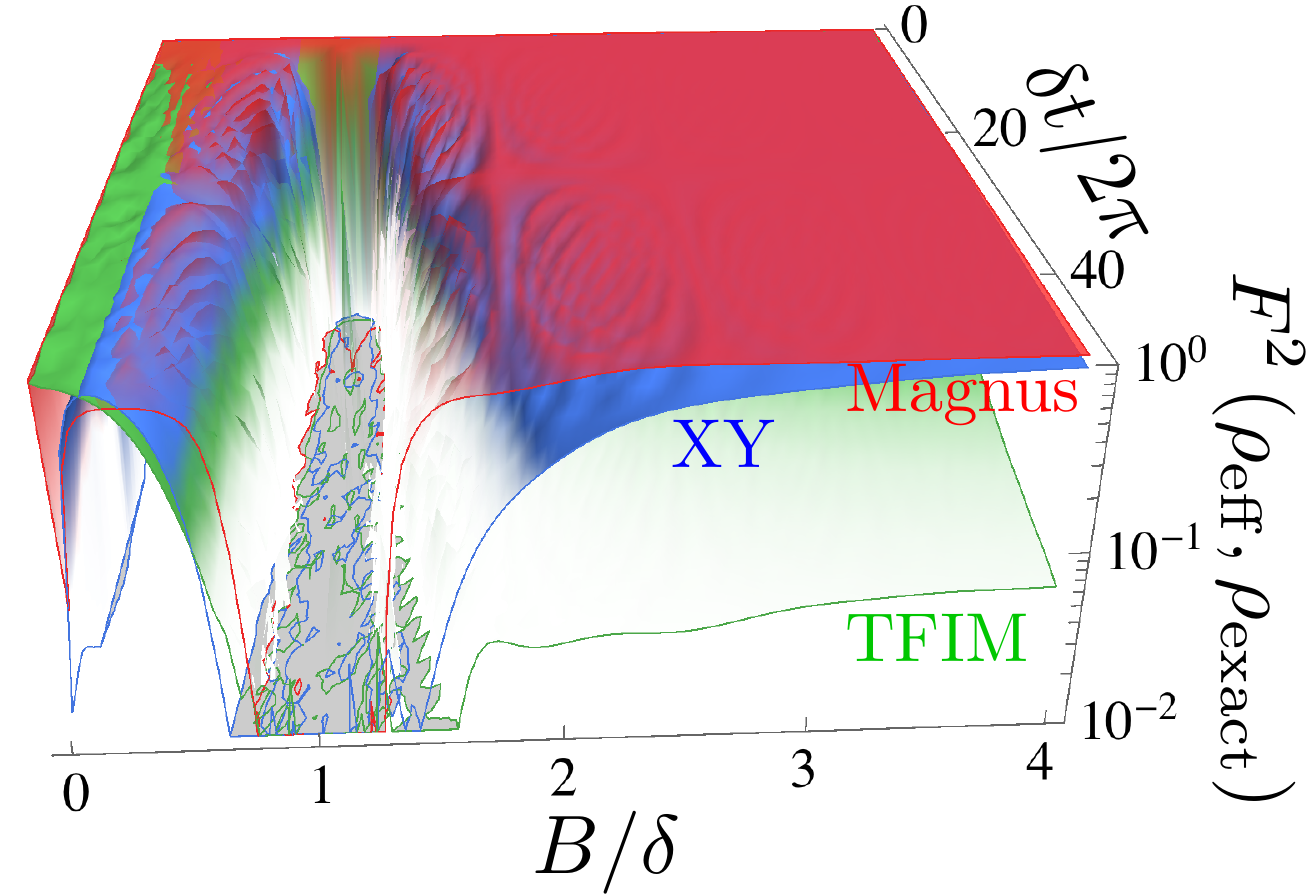}
\caption{(Color online)  \emph{Fidelity of approximate models with exact solution.}  The fidelity of the second-order Magnus series Eq.~\eqref{eq:NPMagnus} (red), the XY model Eq.~\eqref{eq:XYeff} (blue), and the TFIM Eq.~\eqref{eq:TFIM} (green) with the exact numerical dynamics as a function of time and transverse magnetic field strength.  The XY and Magnus descriptions have good fidelity away from the ``resonance" at $B\sim \delta$ and small $B\sim 0$, while the TFIM only has good fidelity for small $B/\delta$.}
\label{fig:gp2Fidelities}
\end{figure}

Our first characterization of the accuracy of the effective models is given by the fidelity $F\left(\rho_{\mathrm{eff}},\rho_{\mathrm{exact}}\right)=\mathrm{Tr}\left[\sqrt{\sqrt{\rho_{\mathrm{eff}}} \rho_{\mathrm{exact}}\sqrt{\rho_{\mathrm{eff}}}}\right]$ as a function of time starting from all spins pointing along the $y$ direction (i.e., perpendicular to both the transverse field and the spin-dependent force) and the vacuum boson state.  For comparisons of the second-order Magnus series Eq.~\eqref{eq:NPMagnus} dynamics with the exact dynamics, $\rho_{\mathrm{eff}}=|\psi_{\mathrm{eff}}\left(t\right)\rangle\langle \psi_{\mathrm{eff}}\left(t\right)|$ and $\rho_{\mathrm{exact}}=|\psi_{\mathrm{exact}}\left(t\right)\rangle\langle \psi_{\mathrm{exact}}\left(t\right)|$ with both $|\psi_{\mathrm{eff}}\left(t\right)\rangle$ and $|\psi_{\mathrm{exact}}\left(t\right)\rangle$ consisting of pure states of spins and bosons, and so $F\left(\rho_{\mathrm{eff}},\rho_{\mathrm{exact}}\right)=\left| \langle \psi_{\mathrm{eff}}\left(t\right)|\psi_{\mathrm{exact}}\left(t\right)\rangle\right|$.  When comparing the spin-only XY and TFIM models with the exact solution, we instead take the spin density matrices $\rho_{\mathrm{eff}}=|\psi_{\mathrm{eff}}\left(t\right)\rangle\langle \psi_{\mathrm{eff}}\left(t\right)|$ and $\rho_{\mathrm{exact}}=\mathrm{Tr}_{\mathrm{bosons}}|\psi_{\mathrm{exact}}\left(t\right)\rangle\langle \psi_{\mathrm{exact}}\left(t\right)|$, in which case $F^2\left(\rho_{\mathrm{eff}},\rho_{\mathrm{exact}}\right)=\langle \psi_{\mathrm{eff}}\left(t\right)|\rho_{\mathrm{exact}}|\psi_{\mathrm{eff}}\left(t\right)\rangle$.

\begin{figure*}
\centering
\includegraphics[width=1.6\columnwidth]{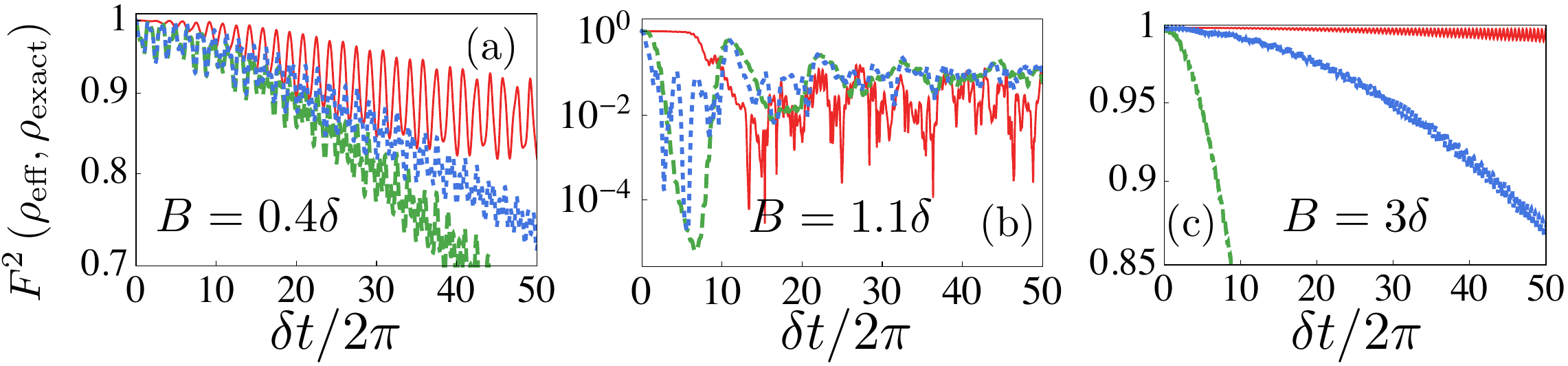}
\caption{(Color online)  \emph{Dynamics of fidelity in various transverse field regimes.}  Comparison of the fidelity of the second-order Magnus Eq.~\eqref{eq:NPMagnus} (red solid), the XY model Eq.~\eqref{eq:XYeff} (blue dotted), and the TFIM Eq.~\eqref{eq:TFIM} (green dashed) with the exact dynamics as a function of time.  In the $B\sim J$ regime (a), the TFIM and XY evolutions are comparable, with the XY having slightly higher fidelity.  In the resonance regime (b), the Magnus series is only accurate at short times, and the strong buildup of spin-boson entanglement precludes any spin-only description.  In the strong-field regime (c), the spin-spin evolution becomes XY-like, with non-perturbatively renormalized spin-spin couplings, and the TFIM description fails.}
\label{fig:DynamicsRegimes}
\end{figure*}

\begin{figure*}
\centering
\includegraphics[width=1.9\columnwidth]{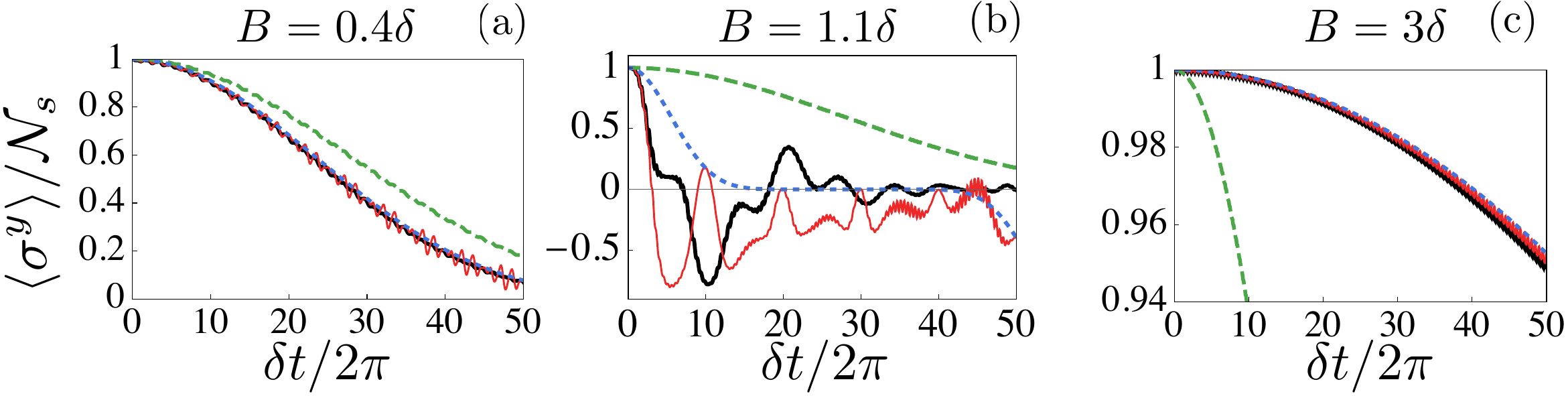}
\caption{(Color online)  \emph{Dynamics of the collective spin.}  Dynamics of the $y$ component of the collective spin predicted by the exact solution (thick black solid), the Magnus series Eq.~\eqref{eq:NPMagnus} (thin red solid), the XY model Eq.~\eqref{eq:XYeff} (blue dotted), and the TFIM Eq.~\eqref{eq:TFIM} (green dashed).  The transverse field rotation has been taken out for clarity (see text for details).  The depolarization of the spin is well captured by either the TFIM or the XY model in the crossover regime (panel (a)).  In the resonant regime (b), strong spin-boson entanglement affects the spin dynamics, and the Magnus series only converges at short times.  In the strong field regime (c), the TFIM misses non-perturbative renormalization of the spin-spin coupling constants, and so fails to capture the timescale of collective demagnetization.}
\label{fig:SpinDynamics}
\end{figure*}

The results for the fidelities at $g/\delta=0.2$ and $\ns=11$ spins are shown in Fig.~\ref{fig:gp2Fidelities} as a function of time and transverse field strength.  A few features are immediately apparent.  First, the approximate second-order Magnus series reproduces the exact dynamics well across the entire range of transverse field except for very small $B$ and a window around $B\sim \delta$.  Additionally, we see that the XY model description performs reasonably well when the second-order Magnus series converges, while the TFIM description only has good fidelity in the small-$B$ region where the Magnus series performs poorly.  Based on this, we can identify five distinct regimes: (1) $B=0$, where the dynamics is known to be an Ising model with spin-dependent boson displacements (see Eq.~\eqref{eq:IsingMagExact}); (2) $0<B\lesssim J$ with $J\sim g^2/\delta$ the spin-spin coupling constants Eq.~\eqref{eq:Jtil}, where the TFIM performs well and the Magnus series generally does not; (3) $J\lesssim B <\delta$ where the Magnus and XY descriptions have reasonable fidelity and the TFIM performance degrades; (4) $B\sim \delta$ where no spin model performs well and the Magnus series does not converge; and (5) $\delta<B$ where the Magnus and XY descriptions perform well and the TFIM description approximates the exact dynamics poorly.  We note that the Magnus series becomes exact when $B=0$, but the XY model derived from it, Eq.~\eqref{eq:XYeff}, performs poorly due to ignoring the $\sin(2Bt)/2B$ terms in the spin-spin couplings Eqs.~\eqref{eq:Jzz}-\eqref{eq:Jyy}.

In Fig.~\ref{fig:DynamicsRegimes} we show a comparison of the dynamics of the fidelities of our approximate approaches with the exact dynamics.  Panel (a) shows the behavior in regime (3), where both the TFIM and XY descriptions have a reasonable fidelity with the XY model performing slightly better.  A more detailed study of this regime, including the crossover from TFIM to XY behavior, will be given in Sec.~\ref{sec:Crossover}.  The dynamics near the resonant regime (4) are shown in panel (b).  Here, we see that neither the spin model descriptions nor the Magnus series has good fidelity at long times, though the Magnus series is reasonable at short times.  The resonant regime is discussed further in Sec.~\ref{sec:resonant}.  Finally, Panel (c) displays the dynamics in the regime (5).  Here, the spin dynamics is governed by an XY model whose spin-spin couplings are non-perturbatively renormalized by the transverse field strength $B$, see Eq.~\eqref{eq:XYeff}.  The TFIM description misses this non-perturbative renormalization, and so fails to provide an accurate description of the spin dynamics.

The fidelity is quite a stringent criterion for comparing how accurate a particular model is for quantum simulation.  Instead, many quantum simulators are focused on measurements of low-order spin correlation functions.  A key observable for trapped ion quantum simulators is the depolarization of the collective spin, which signals the buildup of higher-order spin correlations.  For a collective spin prepared perpendicular to the axis of an Ising coupling without transverse field, there is no mean-field dynamics, and the collective spin only depolarizes due to interactions~\cite{Dylewsky}.  With a transverse field, the collective spin rotates about the field in addition to the interaction-induced depolarization.  All models reproduce the single-particle rotation well, and so for clarity we will remove this trivial rotation by acting on the state with $\exp(i\hat{H}_B t)$.  We show the dynamics of the $y$-component of the collective spin following this rotation in Fig.~\ref{fig:SpinDynamics}, with the thick black solid, thin red solid, blue dotted, and green dashed curves corresponding to the exact, Magnus, XY, and TFIM evolutions.  In the regimes where the Magnus series converges (panels (a) and (c)), the XY model performs well and captures the decay of the magnetization due to coherent spin-spin interactions.  The TFIM, on the other hand, misses the non-perturbative renormalization of the spin-spin coupling constants in the strong field regime, and so fails to predict the correct demagnetization timescale (panel (c)).  In the resonant regime (panel (b)), no spin model correctly reproduces the magnetization dynamics due to strong spin-boson entanglement, and the Magnus series only converges at short times.

\begin{figure}[t]
\centering
\includegraphics[width=0.76\columnwidth]{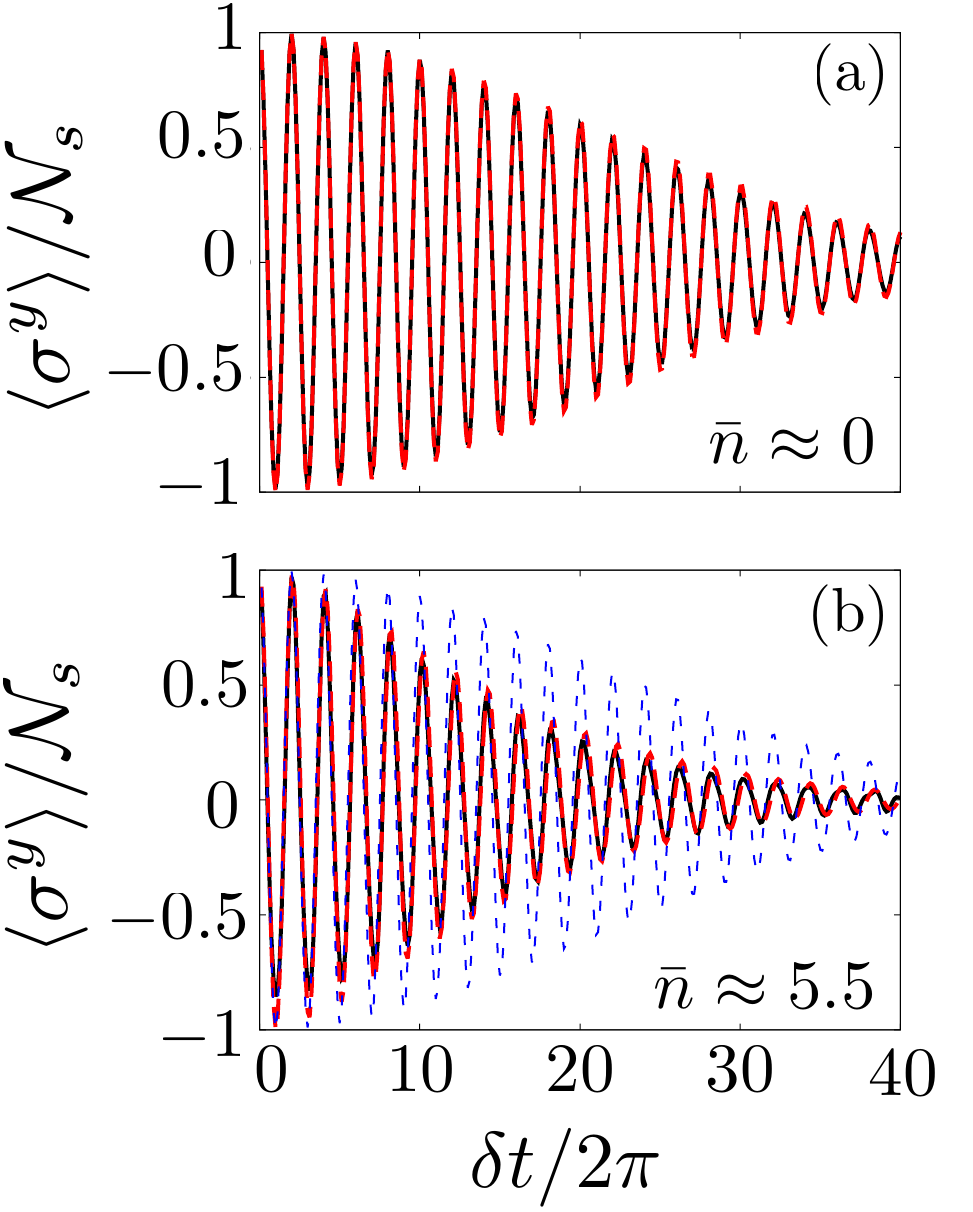}
\caption{(Color online)  \emph{Magnetization dynamics at finite temperature.} Comparison of the exact dynamics (solid black) and dynamics predicted by a incoherent sum of XY spin models' dynamics (red dashed) for the $y$-component of the collective spin at low (panel (a)) and high (panel (b)) temperatures and $B=\delta/4$.  The thin blue dashed line in panel (b) is the zero-temperature XY prediction for comparison, showing that an average of XY models' dynamics captures the thermal dephasing effect well.}
\label{fig:ThermDephasing}
\end{figure}

The effective transverse field that appears in the second-order Magnus series Eq.~\eqref{eq:NPMagnus} is proportional to the thermal energy of the boson modes, and so is expected to give rise to thermal dephasing at nonzero boson temperature $T$ in oscillator units.  Fig.~\ref{fig:ThermDephasing} shows the dynamics of the $y$ component of the collective spin at $B=\delta/4$ and low and high boson temperatures, where boson temperature is expressed in terms of the mean number of quanta $\bar{n}$.  The upper panel compares the dynamics at zero temperature (which is the case also give in Figs.~\ref{fig:gp2Fidelities}-\ref{fig:SpinDynamics}), where the effective spin model is given by Eq.~\eqref{eq:XYeff} with $n=0$ and performs well.  At high temperature, the exact dynamics (black solid) disagrees with the predictions of the model Eq.~\eqref{eq:XYeff} with $n=0$ due to thermal dephasing.  However, taking an incoherent, weighted sum of XY dynamics with different values of $n$ in Eq.~\eqref{eq:XYeff} and Boltzmann weights $\propto e^{-n/T}$, reproduces the thermal dephasing well, as shown by the red dashed curve in Fig.~\ref{fig:ThermDephasing}(b).  We also note that the term causing this thermal dephasing looks like a static field in the secular approximation, and so its dominant effects can be removed by performing a spin echo sequence\footnote{It should be noted that the spin echo pulse does not generically commute with the spin part of the interaction picture rotation operator.  For the single-beam MS implementation, the large separation of timescales between the Magnus and rotation operators makes this fact irrelevant.  For the boson-mediated Ising simulator in a transverse field, this issue can be averted by applying the spin echo pulse at integer multiples of the transverse field rephasing time $2\pi/B$, where the rotating frame rotation operator is proportional to the identity.}.  Finally, we note that this thermal dephasing is expected for the ``imbalanced" MS scheme we propose, as the original balanced MS gate was designed specifically to remove boson mode dependence from the spin dynamics~\cite{PhysRevA.62.022311,PhysRevLett.82.1971}.  However, it is interesting to note that exactly this same thermal dephasing appears when considering the boson-mediated Ising simulator in a transverse field.

\begin{figure}[t]
\centering
\includegraphics[width=\columnwidth]{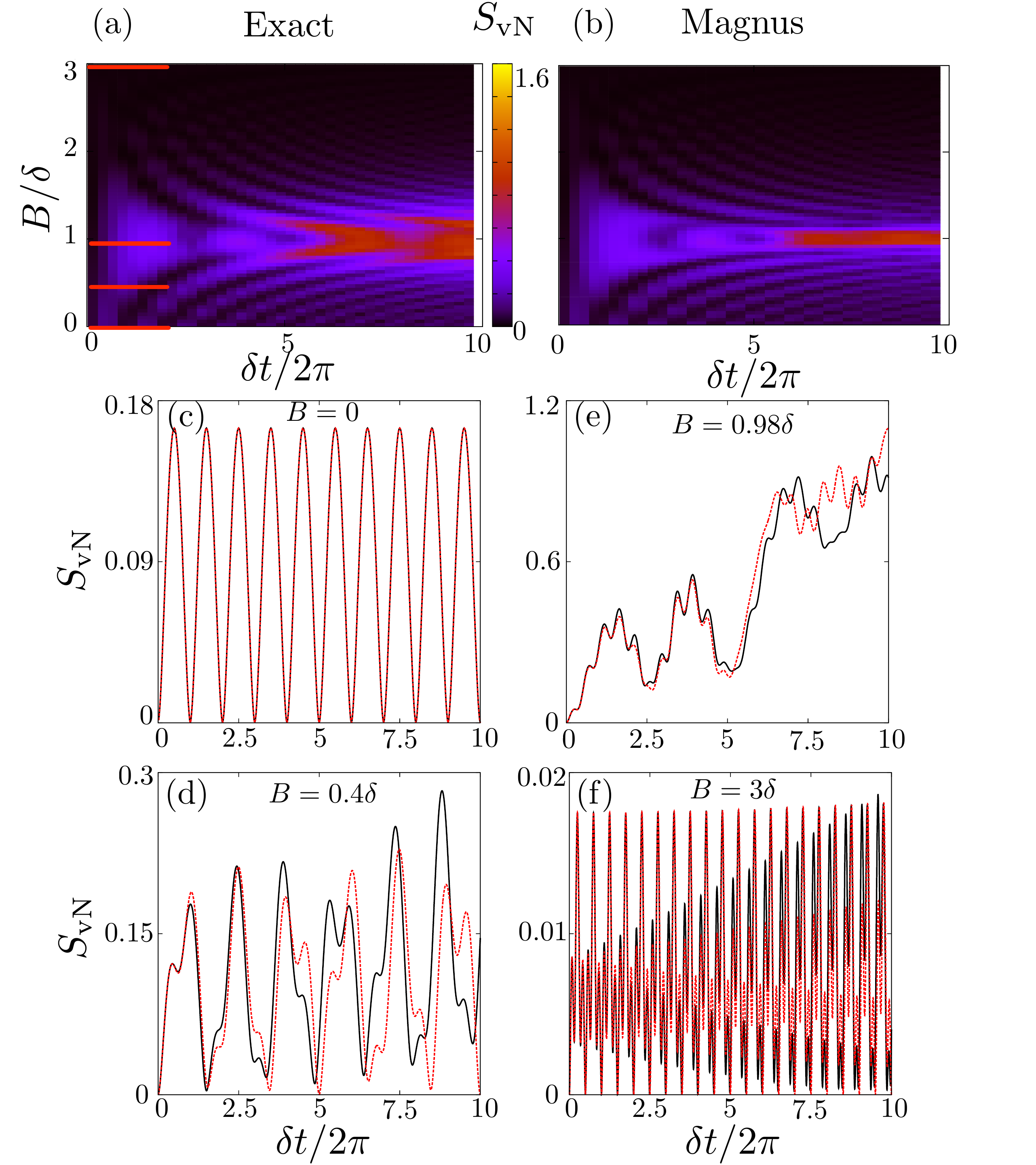}
\caption{(Color online)  \emph{Spin-boson entanglement.} The entanglement between spins and bosons, characterized by the von Neumann entropy of entanglement of the density matrix obtained by tracing out the bosons, as a function of time and transverse field strength is given for the exact dynamics (panel (a)) and the dynamics predicted by the second-order Magnus series Eq.~\eqref{eq:NPMagnus} (panel (b)).  Comparisons of the exact and Magnus dynamics are given for the red lines indicated in panel (a) in panels (c)-(f).  Exact decoupling points are seen in the Ising case (c), but are no longer exact for non-zero transverse field (d).  In the resonant regime (e), strong spin-boson entanglement is present at all times, and well-captured by the Magnus series at short times.  In the strong-field regime, the overall degree of spin-boson entanglement is reduced with respect to the Ising case.}
\label{fig:SBE}
\end{figure}

In general, experiments aiming at simulating the behavior of quantum spin systems would like to minimize the entanglement between the spins and the bosons, as this entanglement leads to a loss of fidelity for the simulation of the pure spin system.  In the Ising case, even when the spin-boson coupling is strong and significant entanglement is built up, there exist certain decoupling times $t_d$ where this entanglement vanishes for a single mode.  Such decoupling points also exist in our approximate Magnus series when $B/\delta$ is a rational fraction, but only up to second order.  In Fig.~\ref{fig:SBE} we show the dynamics of the spin-boson entanglement, characterized by von Neumann entropy of entanglement $S_{\mathrm{vN}}=-\sum_{j}\lambda_{j}\log \lambda_j$, where $\lambda_j$ are the eigenvalues of the reduced density matrix obtained by tracing out the boson modes.  The general structure of the spin-boson entanglement is captured by the second-order Magnus series, as is shown by the comparisons in panels (a) and (b).  However, a detailed analysis (panels (c)-(f)) of the transverse fields marked in panel (a) shows finer-scale structure which is occasionally missed in this approximation.  Panel (c) shows the Ising case in which $B=0$, where the Magnus series is exact and reproduces the bounded, periodic spin-boson entanglement which vanishes at multiples of the decoupling time $t_d$.  In panel (d) we show the spin-boson entanglement for $B=0.4\delta$.  Here, the second-order Magnus series (red dashed) predicts decoupling points at integer multiples of $\tilde{t}_d=10\pi/\delta$, but the exact dynamics (black solid) shows that the spins and bosons do not decouple due to higher order processes.  In the resonant regime (panel (e)), strong spin-boson entanglement is present at all times, and this buildup is reasonably captured at short times by the Magnus series.  Finally, in the strong-field regime (panel(f)), the overall scale of spin-boson entanglement is reduced compared to the Ising case, and is well-captured by the Magnus approximation.  However, the decoupling points $\tilde{t}_d=2\pi/\delta$ predicted by the Magnus series are not exact, and so deviations can be seen from the exact solution at later times.

\section{Failure of the XY model} 
\label{sec:Fail}

The above numerical analysis shows that the approximate second-order Magnus series captures the full dynamics accurately out to experimental timescales except when the transverse field strength is very weak or near the ``resonant" point $B\sim\delta$.  In this section, we look more closely at the regimes where this approach fails, and what the appropriate description of the spin physics is.

\subsection{Resonant regime}
\label{sec:resonant}

For situations in which the coupling $g_{\mu,j}$ is comparable to the resonance parameter $(\delta_{\mu}^2-B^2)$, the Magnus series does not converge beyond short times, as all terms in the infinite-order series contribute strongly at longer timescales.  However, near this point it is useful to re-write the Hamiltonian in the rotating frame of the transverse field, Eq.~\eqref{eq:IPwithB} in terms of $\hat{\sigma}^{\pm}_j=(\hat{\sigma}^z_j\mp i \hat{\sigma}^y)/2$ to find 
\begin{align}
\label{eq:IPRabi} \hat{\mathcal{H}}_I\left(t\right)&=-\sum_{j,\mu}\frac{g_{\mu j}}{2}\left(\hat{a}_{\mu}e^{i\delta_{\mu} t}+\hat{a}_{\mu}^{\dagger}e^{-i\delta_{\mu} t}\right)\left(e^{-i B t}\hat{\sigma}^{+}_j+e^{ i B t}\hat{\sigma}^{-}_j\right)\, .
\end{align}
In particular, if we write $B=\delta_{}+\Delta$, then we have (in the single-mode case)
\begin{align}
\nonumber &\hat{\mathcal{H}}_I\left(t\right)=-\sum_{j}\frac{g_{ j}}{2}\left(\hat{a}_{}e^{i\delta_{} t}+\hat{a}_{}^{\dagger}e^{-i\delta_{} t}\right)\\
&\times \left(e^{-i\delta_{} t-i\Delta t}\hat{\sigma}^{+}_j+e^{i\delta_{} t+\Delta t}\hat{\sigma}^{-}_j\right)\, ,\\
&=\sum_{j}\frac{g_{ j}}{2}\left(e^{i\Delta t} \hat{a}_{}\hat{\sigma}^{+}+e^{i\left(\delta+\Delta\right) t}\hat{a}^{\dagger}_{}\hat{\sigma}^++\mathrm{H.c.}\right)\, .
\end{align}
The terms which rotate as $\Delta$ form the Jaynes-Cummings (JC) model, and the terms with phases $(\delta+\Delta)$ are the counter-rotating terms.  On resonance, $\Delta=0$, the JC terms are responsible for the strong buildup of spin-boson entanglement, and the counter-rotating terms give rise to residual spin-spin interactions.

Taken together, the JC and counter-rotating terms define a multi-spin generalization of the Rabi model~\cite{PhysRev.49.324,Braak}.  For a single boson with uniform coupling to all spins, this is also referred to as the Dicke model~\cite{hepp1973superradiant}.  At resonance, we can tune the parameter regime of the Rabi model from the weak coupling $g\ll \delta$, where cavity QED experiments usually operate to the deep strong coupling regime $g\gtrsim \delta$ which is extremely difficult to access in QED, and has also been challenging to reach in superconducting qubits and other synthetic QED platforms~\cite{PhysRevLett.105.237001,niemczyk2010circuit}.  The physics in these regimes, where keeping only the JC terms is invalid, can be quite different from the physics of the weak-coupling JC model~\cite{PhysRevLett.105.263603,PhysRevX.2.021007}.  We note that many other proposals exist for realizing Rabi or Dicke models in trapped ions, either for a single ion~\cite{pedernales2015quantum,puebla2016robust}, or for many~\cite{PhysRevLett.112.023603}.

It is worth pointing out that in the single-beam MS realization Eq.~\eqref{eq:MSfinaleff}, the resonance that occurs at $B=\delta$ corresponds to driving a sideband of boson motion exactly on resonance, i.e. $\omega_{\mathrm{eff}}=\omega_0$.  This resonance is similar to the one in the boson-mediated Ising simulator when $\delta\to 0$.  The Jaynes-Cummings-dominated physics of this resonance has been used to experimentally generate nonclassical phonon states for a single trapped ion~\cite{PhysRevLett.76.1796}.

\subsection{Weak transverse field:Transverse-field Ising model and XY crossover}
\label{sec:Crossover}

The other place where the XY model description fails is at small transverse fields $B$ relative to the Ising spin-spin coupling constants $J$.  However, this is precisely the regime where one could expect that a perturbative analysis in $\hat{H}_B$ could perform well.  One means to derive a spin model that is perturbative in $\hat{H}_B$ is to generate the Magnus series for the Hamiltonian of the spin-dependent force and transverse field Eq.~\eqref{eq:HI} without transforming to the rotating frame of the transverse field.  The first two orders of this $B$-perturbative Magnus series are
\begin{align}
\label{eq:1p}&\hat{\mathcal{A}}_1^{(p)}\left(t\right)=-i\int_0^tdt_1\left(\hat{{H}}_I\left(t_1\right)+\hat{H}_B\right)\\
&=\sum_{\mu j}g_{\mu j} \left[\alpha_{\mu}\left(t\right)\hat{a}^{\dagger}_{\mu}-\mathrm{H.c.}\right]\hat{\sigma}^z_j+i \frac{B t}{2}\sum_j \hat{\sigma}^x_j\, ,\\
\label{eq:2p}&\hat{\mathcal{A}}_2^{(p)}\left(t\right)=-\frac{1}{2}\int_0^tdt'\left[\hat{\mathcal{A}}_1^{(p)}\left(t'\right),-i\hat{H}_I\left(t\right)\right]\, ,\\
&=\sum_{\mu j}g_{\mu j}\left(\gamma_{\mu}(t) \hat{a}^{\dagger}_{\mu} -\mathrm{H.c.}\right)\hat{\sigma}^y_j-i\sum_{j,j'}\tilde{J}_{j,j'}(t)\hat{\sigma}^z_j\hat{\sigma}^z_{j'}\, ,
\end{align}
where $\alpha_{\mu}(t)$ and $\tilde{J}_{j,j'}(t)$ are the same spin-boson and spin-spin coupling for the $B\to 0$ case, see Eqs.~\eqref{eq:alp} and \eqref{eq:Jtil}, and
\begin{align}
\gamma_{\mu}(t)&=\frac{B}{4\delta_{\mu}^2}\left[\delta_{\mu} t\left(1-e^{-i\delta_{\mu} t}\right)+2i\left(1-e^{-i\delta_{\mu} t}\right)\right]\, .
\end{align}
In addition to the TFIM that might be naively expected, the truncated description also contains spin-dependent displacements which depend upon both the $z$- and $y$-components of the spin.  We see that higher-order terms in the Magnus series which are beyond the TFIM description (namely, the $\gamma_{\mu}$ terms), vanish at the decoupling points $t_d$ defined above, but also have a norm which grows as $Bt$, and so will only be small enough to be considered perturbations for times scaling as $t\ll1/B$.

An alternative approach, which does not suffer from the restrictions on timescales of the $B$-perturbative Magnus series, is to perform a canonical transformation which removes $\hat{H}_{\mathrm{SB};I}(t)$ from the Hamiltonian, as first introduced in the time-independent case by Porras and Cirac~\cite{PhysRevLett.92.207901} and in the time-dependent case by Wang and Freericks~\cite{PhysRevA.86.032329}.  We can alternately view this procedure by transforming to an interaction picture rotating with $\hat{H}_{\mathrm{SB};I}(t)$, but approximating the interaction picture rotation operator by only the first term in the Magnus series generated by $\hat{H}_{\mathrm{SB};I}(t)$.  Since the Magnus series generated by $\hat{H}_{\mathrm{SB};I}(t)$ exactly terminates, we do not have to worry about delicate issues of convergence.  Noting that $\exp(-i\int_0^t dt'\hat{H}_{\mathrm{SB};I}(t'))=\hat{U}_{\mathrm{SB}}\left(t\right)$, the spin-boson propagator from the pure driving case above, and writing the propagator as $\hat{U}(t)=\hat{U}_{\mathrm{SB}}(t)\hat{\mathcal{U}}(t)$, the Schr\"odinger equation becomes
\begin{align}
\nonumber i\partial_t  \hat{\mathcal{U}}(t)=\Big(&\sum_{\mu,j,j'}g_{\mu j}g_{\mu j'}\frac{1-\cos\left(\delta_{\mu} t\right)}{4\delta_{\mu}}\hat{\sigma}^z_{j}\hat{\sigma}^z_{j'}+\hat{H}_B\\
\label{eq:CanonEff}&+\sum_{n=1}^{\infty}\frac{\left(-1\right)^n}{n!}\mathcal{C}^{(n)}\left(\hat{\mathcal{A}}_1^{(p)}(t),\hat{H}_B\right)\Big)\hat{\mathcal{U}}(t)\, ,
\end{align}
in which $\mathcal{C}^{(n)}(\hat{A},\hat{B})$ denotes the $n^{\mathrm{th}}$ nested commutator of $\hat{A}$ with $\hat{B}$, e.g., $\mathcal{C}^{(2)}(\hat{A},\hat{B})=[\hat{A},[\hat{A},\hat{B}]]$.  As expected, we have that
\begin{align}
\sum_{\mu} g_{\mu j}g_{\mu j'}\int_0^{t} dt'\frac{1-\cos\left(\delta_{\mu} t'\right)}{4\delta_{\mu}}&=\tilde{J}_{j,j'}\left(t\right)\, ,
\end{align}
and so this approach reproduces the exactly terminating Magnus series given above when $\hat{H}_B\to 0$.

If we neglect all of the commutators in Eq.~\eqref{eq:CanonEff}, then this defines an effective Hamiltonian which is a time-dependent TFIM.  Due to the time-dependence, operator character beyond just the TFIM can appear in the evolution operator (Appendix \ref{app:Can}).  That is to say, one should be cautious about referring to the effective Hamiltonian $\hat{H}_{\mathrm{eff}}$, even in the absence of boson effects, as corresponding to a TFIM with spin-spin couplings given by $\tilde{J}_{j,j'}(t)$.  In addition, it can be shown (Appendix \ref{app:Can}) that the terms in the ``correction series" given by the commutators in Eq.~\eqref{eq:CanonEff}--even though they appear to involve only spin-boson couplings--lead to effective spin-spin interactions of the same order in $g/\delta$ as the Ising spin-spin interactions.  When $\hat{H}_B$ can no longer be considered a perturbation, these corrections and their virtually mediated spin-spin interactions will become important.

\begin{figure}[t]
\centering
\includegraphics[width=0.66\columnwidth]{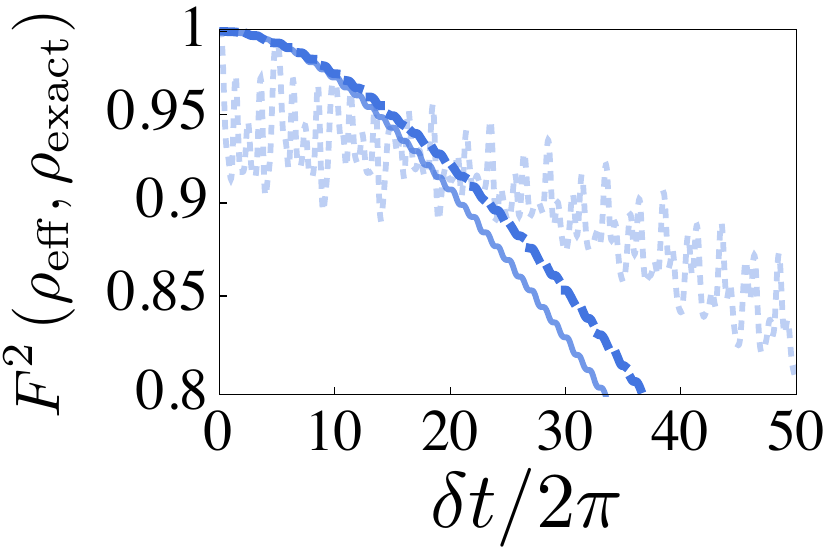}
\caption{(Color online)  \emph{Comparison of orders in the series Eq.~\eqref{eq:CanonEff}.} The fidelity of the zeroth-order (dotted), first-order (solid), and second-order approximate dynamics predicted by the series Eq.~\eqref{eq:CanonEff} with the exact solution at $B=0.4\delta$ are shown as a function of time.  Including more terms in the series always improves the fidelity at short times, but may not improve the fidelity at long times.}
\label{fig:Can}
\end{figure}

A natural question to ask is whether the series Eq.~\eqref{eq:CanonEff} converges, and how it compares with the other approaches considered in this work.  In Fig.~\ref{fig:Can} we show the fidelity of the exact solution with the first three orders of Eq.~\eqref{eq:CanonEff}, corresponding to truncating the correction series at $n=0$, $1$, and $2$, again restricting to the single-mode case with a uniform spin-boson coupling.  The zeroth-order dynamics displays fast oscillations corresponding to neglecting boson displacements proportional to the $y$-component of the spin.  These displacements are captured by the higher-order dynamics, and increasing the order always improves the fidelity at short times.  At longer times, higher-order approximations do not necessarily have a higher fidelity than lower-order approximations.  In part this is due to the explicit time dependence of the terms in the correction series, as processes arising from time-ordering can have similar weight to neglected higher-order terms in the correction series.

\begin{figure}[t]
\centering
\includegraphics[width=0.66\columnwidth]{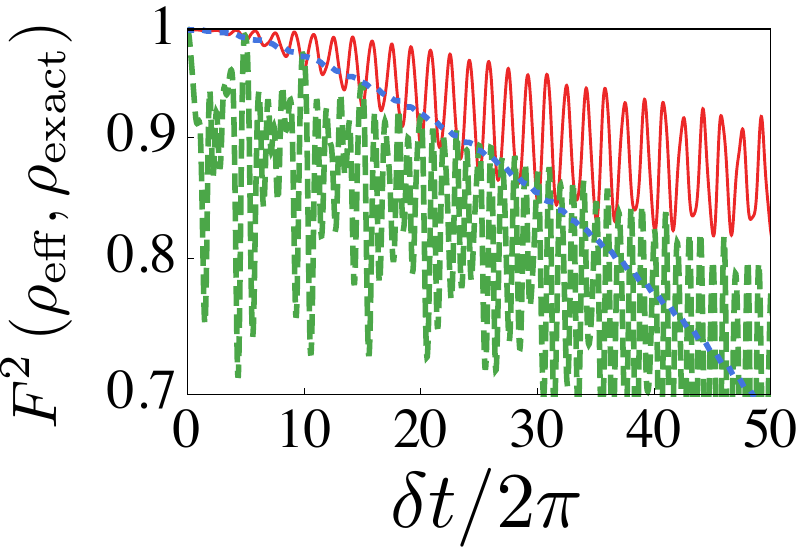}
\caption{(Color online)  \emph{Comparison of Perturbative and Non-perturbative approaches in the crossover regime.} The fidelity of the $B$-non-perturbative Magnus series (red solid), $B$-perturbative Magnus series (green dotted), and second-order canonical transformation (blue dashed) dynamics with the exact dynamics at $B=0.4\delta$ are given as a function of time.  The non-perturbative Magnus series performs best in this regime, and the perturbative Magnus series performs worst.  The canonical transformation dynamics is the same as the blue dashed curve in Fig.~\ref{fig:Can}.}
\label{fig:MagCan}
\end{figure}

We next address how the approaches which are perturbative in $\hat{H}_B$ compare with the non-perturbative Magnus series of Eq.~\eqref{eq:NPMagnus} via the fidelity in Fig.~\ref{fig:MagCan}.  The red solid line is the $B$-non-perturbative Magnus series, the green dotted line is the $B$-perturbative Magnus series Eqs.~\eqref{eq:1p}-\eqref{eq:2p}, and and the blue dashed line is the second-order canonical transformation Eq.~\eqref{eq:CanonEff}.  Here, $B=0.4\delta$, and so this dynamics is in the crossover regime where both the TFIM and XY descriptions perform reasonably.  We see that the non-perturbative Magnus series generally performs the best, with the perturbative Magnus series generally performing the worst (worse even than the zeroth-order canonical transformation, compare Fig.~\ref{fig:Can}).  However, interestingly, the canonical transformation result, which contains terms that are higher-order in the spin-boson coupling and transverse field than the perturbative Magnus series, performs worse than the perturbative Magnus series at later times.

\begin{figure}[t]
\centering
\includegraphics[width=0.76\columnwidth]{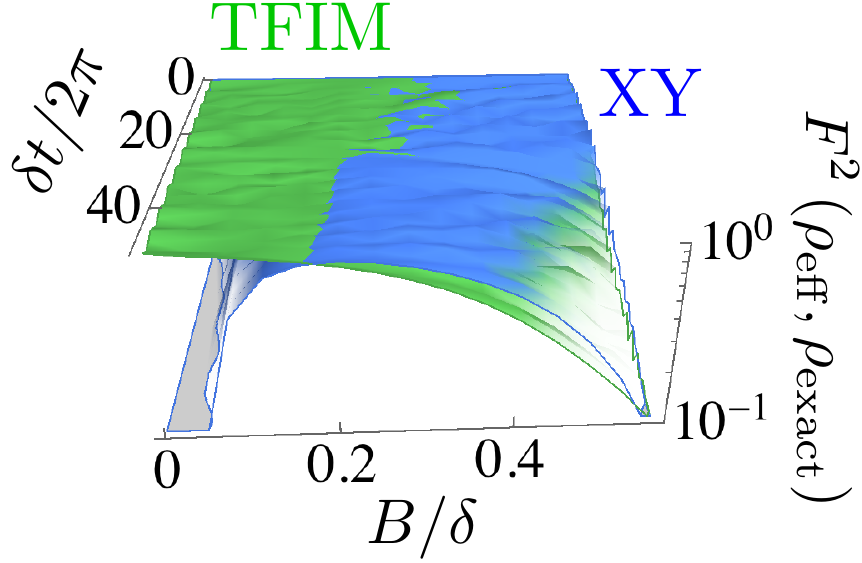}
\caption{(Color online)  \emph{Crossover from TFIM to XY behavior.} The fidelity of the XY model (blue) and TFIM (blue) predictions with the exact solution in the small $B$ regime.  When $B\gtrsim J$ and $B\ll \delta$, both models predict spin-spin couplings of XY character with strengths given by half of the Ising couplings at $B=0$.  In this region the most accurate spin model shifts from being the $B$-perturbative TFIM to being the $B$-non-perturbative XY model.}
\label{fig:Crossover}
\end{figure}

Finally, we discuss the crossover from TFIM-like spin physics to XY-like spin physics as a function of transverse field.  This crossover can be derived starting from the assumed validity of the TFIM,
\begin{align}
\hat{H}_{\mathrm{TFIM}}&=\sum_{j,<j'}J_{j,j'}\hat{\sigma}^z_j\hat{\sigma}^z_{j'}-\frac{B}{2}\sum_{j}\hat{\sigma}^x_j\,.
\end{align}
The Ising coupling can be written in terms of the operators $\hat{\sigma}^{x\pm}_j=\frac{1}{2}\left(\hat{\sigma}^z_j\mp i\hat{\sigma}^y_j\right)$ which create excitations along the field direction as
\begin{align}
\sum_{j,<j'}J_{j,j'}\hat{\sigma}^z_j\hat{\sigma}^z_{j'}&=\sum_{j,<j'}J_{j,j'}\left(\hat{\sigma}^{x+}_j+\hat{\sigma}^{x-}_j\right)\left(\hat{\sigma}^{x+}_{j'}+\hat{\sigma}^{x-}_{j'}\right)\, .
\end{align}
When the field is strong, we expect that the only terms energetically allowed are those that preserve the number of excitations along the field direction, and so we ignore products of two $\hat{\sigma}^{x+}$s or $\hat{\sigma}^{x-}$s, giving
\begin{align}
\sum_{j,<j'}J_{j,j'}\hat{\sigma}^z_j\hat{\sigma}^z_{j'}&\approx \sum_{j,<j'}J_{j,j'}\left(\hat{\sigma}^{x+}_j\hat{\sigma}^{x-}_{j'}+\hat{\sigma}^{x-}_j\hat{\sigma}^{x+}_{j'}+\right)\, ,\\
&=\frac{1}{2}\sum_{j,<j'}J_{j,j'}\left(\hat{\sigma}^z_j\hat{\sigma}^z_{j'}+\hat{\sigma}^y_j\hat{\sigma}^y_{j'}\right)\, .
\end{align}
which is an XY model in the directions perpendicular to the transverse field with spin-spin coupling constants given by half the Ising coupling constants~\cite{richerme2014non,jurcevic2014quasiparticle}.  Noting that $\hat{H}_B$ commutes with these spin-spin interactions, we can alternatively view this as an XY model in a frame rotating with the transverse field, up to corrections scaling as $1/B$.  These are exactly the same spin-spin interactions predicted by the effective XY model Eq.~\eqref{eq:XYeff} in the limit that $B\ll \delta$\footnote{The effective transverse field in \eqref{eq:XYeff} arises due to bosons, and so can not come about starting from an assumed TFIM description.  However, in the limit $B\ll \delta$ where the crossover occurs, it is small and can be neglected.}.  The above analysis suggests that the crossover from TFIM to XY behavior occurs in the regime $B\gtrsim J$ but $B\ll \delta$ where both models predict the same spin-spin coupling physics, and numerical analysis confirms this picture (Fig.~\ref{fig:Crossover}).

\section{Conclusions}
\label{sec:Concl}

In summary, we analyzed the dynamics of a boson-mediated Ising quantum spin simulator in the presence of an effective transverse field of strength $B$ that does not commute with the spin-boson coupling, and identified regimes in which the spin dynamics are captured by pure spin models.  For small transverse field on the order of the Ising spin-spin coupling constants, the dynamics are well-described by the transverse-field Ising model (TFIM), while for larger fields the dynamics has the character of an XY model with possible non-perturbative renormalization of the spin-spin couplings from their Ising values.  While for moderate fields the XY description coincides with the strong-field limit of the TFIM, the XY model becomes the more fundamental description at stronger field, as evidenced e.g.~by a larger fidelity with respect to the true dynamics.  In contrast to the case of pure driving (no transverse field) where spins and bosons stroboscopically decouple from each other, we show that the non-commutativity of the transverse field and spin-boson coupling causes spin-boson entanglement which does not strictly vanish at any time, but can be made parametrically small for experimentally relevant timescales in certain limits.  Our emergent XY model performs well except near $B\sim \delta$, where spin rotation resonantly drives boson excitations and no pure spin description performs well.  In addition, our approach identifies that the XY description also contains an effective transverse field whose strength depends on the boson mode energy, and that thermal dephasing can be captured by considering an incoherent, thermally-weighted sum of XY model dynamics.  We substantiated our analysis with analytical calculations based on truncated Magnus series and numerical calculations.  In addition, we showed that this same analysis applies to a single-beam M{\o}lmer-S{\o}rensen scheme corresponding to off-resonant driving of single sideband of boson excitation in an appropriate rotating frame.

\begin{acknowledgments} 
We would like to acknowledge useful discussions with John Bollinger, Justin Bohnet, and Michael Foss-Feig, and support from NSF-PHY 1521080, JILA-NSF-PFC-1125844, ARO, MURI-AFOSR and AFOSR.  MLW thanks the NRC postdoctoral program for support.
\end{acknowledgments} 

\bibliography{references}

\clearpage 
\onecolumngrid

\appendix
\section{Derivation of the non-perturbative Magnus series to second order}
\label{app:Magnus}

In this appendix, we detail the derivation of the first two orders in the Magnus series of the interaction picture Hamiltonian
\begin{align}
\hat{\mathcal{H}}_I\left(t\right)&=-\frac{1}{2}\sum_{\mu j} g_{\mu j}\left(\hat{a}_{\mu} e^{i\delta_{\mu} t}+\hat{a}_{\mu}^{\dagger}e^{-i\delta_{\mu} t}\right)\left(\cos\left(Bt\right)\hat{\sigma}^z_j-\sin\left(B t\right)\hat{\sigma}^y_j\right)\, .
\end{align}
The first-order term is
\begin{align}
\hat{\mathcal{A}}_1\left(t\right)&=-i\int_0^tdt_1\hat{\mathcal{H}}_I\left(t_1\right)=\sum_{\mu j}g_{\mu j} \left[\alpha^z_{\mu}\left(t\right)\hat{a}^{\dagger}_{\mu}\hat{\sigma}^z_j+\alpha^y_{\mu}\left(t\right)\hat{a}_{\mu}^{\dagger}\hat{\sigma}^y_j-\mathrm{H.c.}\right]\, ,
\end{align}
where
\begin{align}
\alpha^z_{\mu}\left(t\right)&=\frac{i}{2}\int_0^{t}dt_1 e^{-i\delta_{\mu} t_1}\cos\left( Bt_1\right)=\frac{1}{2} \frac{\delta_{\mu} -e^{-i\delta_{\mu} t}\left(\delta_{\mu} \cos\left(B t\right)+ iB \sin\left( B t\right)\right)}{\delta_{\mu}^2-B^2}\, ,\\
\alpha^y_{\mu}\left(t\right)&=-\frac{i}{2}\int_0^{t}dt_1 e^{-i\delta_{\mu} t_1}\sin\left( Bt_1\right)=\frac{1}{2} \frac{i B +e^{-i\delta_{\mu} t}\left(\delta_{\mu} \sin\left(B t\right)- iB \cos\left( B t\right)\right)}{\delta_{\mu}^2-B^2}\, .
\end{align}
Using the recursion 
\begin{align}
\label{eq:MagRecur} \hat{\mathcal{A}}_n&=\sum_{j=1}^{n-1}\frac{B_j}{j!} \sum_{\begin{array}{c} k_1+\dots+k_j=n-1\\ k_1\ge 1,\dots,k_j\ge 1\end{array}}\int_0^t dt'\left[\hat{\mathcal{A}}_{k_1}\left(t'\right),\left[\hat{\mathcal{A}}_{k_2}\left(t'\right),\dots ,\left[\hat{\mathcal{A}}_{k_j}\left(t'\right),-i\hat{\mathcal{H}}_I\left(t'\right)\right]\dots \right]\right]\, ,
\end{align}
 the second-order term is $\hat{\mathcal{A}}_2\left(t\right)=-\frac{1}{2}\int_0^tdt'\check{\mathcal{A}}_2\left(t'\right)$, where for future convenience we have defined $\check{\mathcal{A}}_2\left(t\right)=\left[\hat{\mathcal{A}}_1\left(t\right),-i\hat{\mathcal{H}}_I\left(t\right)\right]$.  $\check{\mathcal{A}}_2$ may be written as
\begin{align}
&\check{\mathcal{A}}_2\left(t\right)=\frac{i}{2}\sum_{\mu\mu'}\sum_{jj'}g_{\mu' j'}g_{\mu j}\Big[\left(\alpha^z_{\mu}\left(t\right)\hat{a}^{\dagger}_{\mu}-\bar{\alpha}^z_{\mu}\left(t\right)\hat{a}^{\dagger}_{\mu}\right)\hat{\sigma}^z_j+\left(\alpha^y_{\mu}\left(t\right)\hat{a}^{\dagger}_{\mu}-\bar{\alpha}^y_{\mu }\left(t\right)\hat{a}^{\dagger}_{\mu}\right)\hat{\sigma}^y_j,\\
&\left(e^{i\delta_{\mu'} t}\hat{a}_{\mu'}+e^{-i\delta_{\mu'} t}\hat{a}^{\dagger}_{\mu'}\right)\left(\cos\left(B t\right)\hat{\sigma}^z_{j'}-\sin\left(B t\right)\hat{\sigma}^y_{j'}\right)\Big]\, .
\end{align}
Using the commutator identity $\left[AB,CD\right]=AC\left[B,D\right]+\left[A,C\right]DB$, valid when $A$ commutes with $D$ and $B$ commutes with $C$, we find
\begin{align}
&\check{\mathcal{A}}_2\left(t\right)=\frac{i}{2}\sum_{\mu\mu'}\sum_{jj'}g_{\mu j}g_{\mu' j'}\Big\{\left(\alpha^z_{\mu}\left(t\right)\hat{a}^{\dagger}_{\mu}-\bar{\alpha}^z_{\mu }\left(t\right)\hat{a}_{\mu}\right)\left(e^{i\delta_{\mu'} t}\hat{a}_{\mu'}+e^{-i\delta_{\mu'} t}\hat{a}^{\dagger}_{\mu'}\right)\left[\hat{\sigma}^z_j,\left(\cos\left(B t\right)\hat{\sigma}^z_{j'}-\sin\left(B t\right)\hat{\sigma}^y_{j'}\right)\right]\\
\nonumber &+\left(\alpha^y_{\mu}\left(t\right)\hat{a}^{\dagger}_{\mu}-\bar{\alpha}^y_{\mu}\left(t\right)\hat{a}_{\mu}\right)\left(e^{i\delta_{\mu'} t}\hat{a}_{\mu'}+e^{-i\delta_{\mu'} t}\hat{a}^{\dagger}_{\mu'}\right)\left[\hat{\sigma}^y_j,\left(\cos\left(B t\right)\hat{\sigma}^z_{j'}-\sin\left(B t\right)\hat{\sigma}^y_{j'}\right)\right]\\
\nonumber&+\left[\left(\alpha^z_{\mu}\left(t\right)\hat{a}^{\dagger}_{\mu}-\bar{\alpha}^z_{\mu}\left(t\right)\hat{a}_{\mu}\right),\left(e^{i\delta_{\mu'} t}\hat{a}_{\mu'}+e^{-i\delta_{\mu'} t}\hat{a}^{\dagger}_{\mu'}\right)\right]\left(\cos\left(B t\right)\hat{\sigma}^z_{j'}-\sin\left(B t\right)\hat{\sigma}^y_{j'}\right)\hat{\sigma}^z_j\\
\nonumber&+\left[\left(\alpha^y_{\mu}\left(t\right)\hat{a}^{\dagger}_{\mu}-\bar{\alpha}^y_{\mu}\left(t\right)\hat{a}_{\mu}\right),\left(e^{i\delta_{\mu'} t}\hat{a}_{\mu'}+e^{-i\delta_{\mu'} t}\hat{a}^{\dagger}_{\mu'}\right)\right]\left(\cos\left(B t\right)\hat{\sigma}^z_{j'}-\sin\left(B t\right)\hat{\sigma}^y_{j'}\right)\hat{\sigma}^y_j
\Big\}\, ,
\end{align}
which, upon collecting terms and using commutation relations, gives
\begin{align}
 \nonumber &\check{\mathcal{A}}_2\left(t\right)=\sum_{\mu\mu'}\sum_jg_{\mu j}g_{\mu' j}\hat{\sigma}^x_j \Big[\hat{a}^{\dagger}_{\mu}\hat{a}^{\dagger}_{\mu'}\check{\alpha}^{++}_{\mu\mu'}\left(t\right)-\hat{a}_{\mu}\hat{a}_{\mu'}\bar{\check{\alpha}}^{++}_{\mu\mu'}\left(t\right)+\left(\hat{a}^{\dagger}_{\mu} \hat{a}_{\mu'}\check{\alpha}^{+-}_{\mu\mu'}\left(t\right)-\hat{a}_{\mu}\hat{a}_{\mu'}^{\dagger}\bar{\check{\alpha}}^{+-}_{\mu\mu'}\left(t\right)\right)\left(1-\delta_{\mu,\mu'}\right)\Big]\\
\label{eq:Om2} &-i\sum_{j,j'} \left[\check{J}^{zz}_{j,j'}\left(t\right)\hat{\sigma}^z_{j}\hat{\sigma}^z_{j'}+\check{J}_{j,j'}^{yy}\left(t\right)\hat{\sigma}^y_{j}\hat{\sigma}^y_{j'}-\left(1-\delta_{j,j'}\right)\check{J}^{zy}_{j,j'}\left(t\right)\hat{\sigma}^z_j\hat{\sigma}^{y}_{j'}\right]-i\sum_{\mu}\sum_jg_{\mu j}^2\check{B}_{\mathrm{eff};\mu}\hat{\sigma}^x_j\left(2\hat{a}^{\dagger}_{\mu} \hat{a}_{\mu}+1\right)
\end{align}
Here, we have defined
\begin{align}
\beta_{\mu}\left(t\right)&=\left(\alpha^z_{\mu}\left(t\right)\sin\left(B t\right)+\alpha^y_{\mu}\left(t\right)\cos\left(B t\right)\right)\, ,\\
\check{\alpha}^{+\pm}_{\mu\mu'}\left(t\right)&=-\beta_{\mu}\left(t\right)e^{\mp i\delta_{\mu'} t}=-\left(\alpha^z_{\mu}\left(t\right)\sin\left(B t\right)+\alpha^y_{\mu}\left(t\right)\cos\left(B t\right)\right)e^{\mp i\delta_{\mu'} t}\, ,\\
\check{J}^{zz}_{j,j'}\left(t\right)&=\sum_{\mu}\frac{g_{\mu j'}g_{\mu j}}{2}\left(\bar{\alpha}^z_{\mu }\left(t\right)e^{-i\delta_{\mu} t}+{\alpha}^z_{\mu }e^{i\delta_{\mu} t}\right)\cos\left(B t\right)\, ,\\
\check{J}^{yy}_{j,j'}\left(t\right)&=-\sum_{\mu}\frac{g_{\mu j'}g_{\mu j}}{2}\left(\bar{\alpha}^y_{\mu }\left(t\right)e^{-i\delta_{\mu} t}+{\alpha}^y_{\mu }e^{i\delta_{\mu} t}\right)\sin\left(B t\right)\, ,\\
\check{J}^{zy}_{j,j'}\left(t\right)&=\sum_{\mu}\frac{g_{\mu j'}g_{\mu j}}{2}\left[\left(\bar{\alpha}^y_{\mu }\left(t\right)e^{-i\delta_{\mu} t}+{\alpha}^y_{\mu }e^{i\delta_{\mu} t}\right)\cos\left(B t\right)-\left(\bar{\alpha}^z_{\mu }\left(t\right)e^{-i\delta_{\mu} t}+{\alpha}^z_{\mu }e^{i\delta_{\mu} t}\right)\sin\left(Bt\right)\right]\, ,\\
\check{B}_{\mathrm{eff};\mu}&=-\frac{i}{2}\left(\beta_{\mu}\left(t\right)e^{i\delta_{\mu} t}-\bar{\beta}_{\mu}\left(t\right)e^{-i\delta_{\mu'} t}\right)\, .
\end{align}
With this, we have that the second-order term in the Magnus expansion is
\begin{align}
\hat{\mathcal{A}}_2&=\sum_{\mu\mu'}\sum_jg_{\mu j}g_{\mu' j}\hat{\sigma}^x_j \Big[\hat{a}^{\dagger}_{\mu}\hat{a}^{\dagger}_{\mu'}{\alpha}^{++}_{\mu\mu'}\left(t\right)-\hat{a}_{\mu}\hat{a}_{\mu'}\bar{{\alpha}}^{++}_{\mu\mu'}\left(t\right)+\left(\hat{a}^{\dagger}_{\mu} \hat{a}_{\mu'}{\alpha}^{+-}_{\mu\mu'}\left(t\right)-\hat{a}_{\mu}\hat{a}_{\mu'}^{\dagger}\bar{{\alpha}}^{+-}_{\mu\mu'}\left(t\right)\right)\left(1-\delta_{\mu,\mu'}\right)\Big]\\
 &-i\sum_{j,j'} \left[\tilde{J}^{zz}_{j,j'}\left(t\right)\hat{\sigma}^z_{j}\hat{\sigma}^z_{j'}+\tilde{J}_{j,j'}^{yy}\left(t\right)\hat{\sigma}^y_{j}\hat{\sigma}^y_{j'}+\left(1-\delta_{j,j'}\right)\tilde{J}^{zy}_{j,j'}\left(t\right)\hat{\sigma}^z_j\hat{\sigma}^{y}_{j'}\right]-i\sum_{\mu}\sum_jg_{\mu j}^2{B}_{\mathrm{eff};\mu}\hat{\sigma}^x_j\left(2\hat{a}^{\dagger}_{\mu} \hat{a}_{\mu}+1\right)
\end{align}
where $O\left(t\right)=-\frac{1}{2}\int_0^{t} dt' \tilde{O}\left(t'\right)$ and $\tilde{J}^{\mu \nu}\left(t\right)=-\frac{1}{2}\int_0^{t} dt' \check{J}^{\mu\nu}\left(t'\right)$.  Explicitly evaluating the integrals, we find
\begin{align}
\tilde{J}^{z,z}_{j,j'}&=\sum_{\mu} \frac{g_{\mu,j}g_{\mu,j'}}{8}\delta_{\mu}\left(\frac{t+ t\, \sinc(2Bt)}{\delta_{\mu}^2-B^2}+2\frac{B\cos(\delta_{\mu} t) \sin(Bt)-\delta_{\mu}\cos(Bt)\sin(\delta_{\mu} t)}{(\delta_{\mu}^2-B^2)^2}\right)\, ,\\
\tilde{J}^{z,y}_{j,j'}&=\sum_{\mu}\frac{g_{\mu j}g_{\mu j'}}{4}\frac{t\, \sinc\left(Bt\right)\left(B\sin\left(\delta_{\mu} t\right)-\delta_{\mu} \sin\left(Bt\right)\right)}{\delta_{\mu} ^2-B^2}\, ,\\
\tilde{J}^{y,y}_{j,j'}&=\sum_{\mu}\frac{g_{\mu j}g_{\mu j'} }{8}\left(\frac{\delta_{\mu}(t- t\, \sinc(2Bt))}{\delta_{\mu}^2-B^2} +2B\frac{\delta_{\mu}\sin(Bt)\cos(\delta_{\mu} t)  -B\sin(\delta_{\mu} t) \cos(Bt)}{(\delta_{\mu}^2-B^2)^2}\right)\, ,\\
\alpha^{++}_{\mu,\mu}&=\frac{i}{4} \frac{e^{-i\delta_{\mu} t}\left(\delta_{\mu}\sin\left( Bt\right)-B\sin\left(\delta_{\mu} t\right)\right)}{\delta_{\mu}(\delta_{\mu}^2-B^2)}\, ,\\
B_{\mathrm{eff};\mu}&=\frac{B t}{4\left(\delta_{\mu}^2-B^2\right)}+\frac{\left(B^2+\delta_{\mu}^2\right)\cos\left(\delta_{\mu} t\right)\sin\left(B t\right)-2B\delta_{\mu} \cos\left(B t\right)\sin\left(\delta_{\mu} t\right)}{4\left(\delta_{\mu}^2-B^2\right)^2}\, .
\end{align}

\section{Third-order terms}
\label{app:MagnusTO}

The recursion relation for third order reads
\begin{align}
\hat{\mathcal{A}}_n&=\frac{B_1}{1!}\int_0^t dt'\left[\hat{\mathcal{A}}_{2}\left(t'\right),-i\hat{\mathcal{H}}_I\left(t'\right)\right]+\frac{B_2}{2!}\int_0^t dt'\left[\hat{\mathcal{A}}_1\left(t'\right),\left[\hat{\mathcal{A}}_{1}\left(t'\right),-i\hat{\mathcal{H}}_I\left(t'\right)\right]\right]\\
&=-\frac{1}{2}\int_0^t dt'\left[\hat{\mathcal{A}}_{2}\left(t'\right),-i\hat{\mathcal{H}}_I\left(t'\right)\right]+\frac{1}{12}\int_0^t dt'\left[\hat{\mathcal{A}}_1\left(t'\right),\check{\mathcal{A}}_2\left(t'\right)\right]\\
&=\frac{1}{2}\int_0^t dt'\left(\left[-i\hat{\mathcal{H}}_I\left(t'\right),\hat{\mathcal{A}}_{2}\left(t'\right)\right]+\frac{1}{6}\left[\hat{\mathcal{A}}_1\left(t'\right),\check{\mathcal{A}}_2\left(t'\right)\right]\right)
\end{align}
Both commutators take the form
\begin{align}
&\sum_{\mu j}g_{\mu j}\Big[c_{\mu}\hat{a}^{\dagger}_{\mu}\hat{\sigma}^z_j-\bar{c}_{\mu}\hat{a}_{\mu}\hat{\sigma}^z_j+d_{\mu}\hat{a}^{\dagger}_{\mu}\hat{\sigma}^y_j-\bar{d}_{\mu}\hat{a}_{\mu}\hat{\sigma}^y_j,-i\sum_{j'\ne j''} \left[\tilde{J}^{zz}_{j','j'}\left(t\right)\hat{\sigma}^z_{j}\hat{\sigma}^z_{j''}+\tilde{J}_{j','j'}^{yy}\left(t\right)\hat{\sigma}^y_{j'}\hat{\sigma}^y_{j''}+\tilde{J}^{zy}_{j',j''}\left(t\right)\hat{\sigma}^z_{j'}\hat{\sigma}^{y}_{j''}\right]\\
\nonumber  &+\sum_{\mu'\mu''}\sum_{j'}g_{\mu' j'}g_{\mu'' j'}\hat{\sigma}^x_{j'} \Big[\hat{a}^{\dagger}_{\mu'}\hat{a}^{\dagger}_{\mu''}{\alpha}^{++}_{\mu'\mu''}\left(t\right)-\hat{a}_{\mu'}\hat{a}_{\mu''}\bar{{\alpha}}^{++}_{\mu'\mu''}\left(t\right)+\left(\hat{a}^{\dagger}_{\mu'} \hat{a}_{\mu''}{\alpha}^{+-}_{\mu'\mu''}\left(t\right)-\hat{a}_{\mu'}\hat{a}_{\mu''}^{\dagger}\bar{{\alpha}}^{+-}_{\mu'\mu''}\left(t\right)\right)\left(1-\delta_{\mu',\mu''}\right)\Big]\\
\nonumber &-i\sum_{\mu'}\sum_{j'}g_{\mu' j'}^2{B}_{\mathrm{eff};\mu'}\hat{\sigma}^x_{j'}\left(2\hat{a}^{\dagger}_{\mu'} \hat{a}_{\mu'}+1\right)\Big]\, .
\end{align}
We will break this into three pieces, and further specialize to the single-mode case, by defining
\begin{align}
\mathbb{I}&=-i\sum_{j'\ne j''} \left[\tilde{J}^{zz}_{j','j'}\left(t\right)\hat{\sigma}^z_{j}\hat{\sigma}^z_{j''}+\tilde{J}_{j','j'}^{yy}\left(t\right)\hat{\sigma}^y_{j'}\hat{\sigma}^y_{j''}+\tilde{J}^{zy}_{j',j''}\left(t\right)\hat{\sigma}^z_{j'}\hat{\sigma}^{y}_{j''}\right]\\
\mathbb{II}&=\sum_{j'}g_{\mu j'}g_{\mu j'}^2 \Big[(\hat{a}^{\dagger}_{\mu})^2{\alpha}^{++}_{\mu\mu}\left(t\right)-(\hat{a}_{\mu})^2\bar{{\alpha}}^{++}_{\mu\mu}\left(t\right)\Big]\\
\mathbb{III}&=-i\sum_{\mu}\sum_{j'}g_{\mu j'}^2{B}_{\mathrm{eff};\mu}\hat{\sigma}^x_{j'}\left(2\hat{n}_{\mu} +1\right)\, .
\end{align}
In this notation, we have
\begin{align}
\nonumber &\sum_jg_{\mu j}\Big[c_{\mu}\hat{a}^{\dagger}_{\mu}\hat{\sigma}^z_j-\bar{c}_{\mu}\hat{a}_{\mu}\hat{\sigma}^z_j+d_{\mu}\hat{a}^{\dagger}_{\mu}\hat{\sigma}^y_j-\bar{d}_{\mu}\hat{a}_{\mu}\hat{\sigma}^y_j,\mathbb{I}\Big]=2\sum_{j\ne j'}g_{\mu j}\Big\{\hat{\sigma}^x_{j}\hat{\sigma}^z_{j'}\left[2\tilde{J}_{j,j'}^{zz}\left(t\right)\left(d_{\mu}\hat{a}^{\dagger}_{\mu}-\bar{d}_{\mu}\hat{a}_{\mu}\right)-\tilde{J}^{zy}_{j,j'}\left(t\right)\left(c_{\mu}\hat{a}^{\dagger}_{\mu}-\bar{c}_{\mu}\hat{a}_{\mu}\right)\right]\\
&-\hat{\sigma}^x_{j}\hat{\sigma}^y_{j'}\left[2\tilde{J}_{j,j'}^{yy}\left(t\right)\left(c_{\mu}\hat{a}^{\dagger}_{\mu}-\bar{c}_{\mu}\hat{a}_{\mu}\right)-\tilde{J}^{zy}_{j,j'}\left(t\right)\left(d_{\mu}\hat{a}^{\dagger}_{\mu}-\bar{d}_{\mu}\hat{a}_{\mu}\right)\right]\Big\}\, ,\\
\nonumber &\sum_{ j}g_{\mu j}\Big[c_{\mu}\hat{a}^{\dagger}_{\mu}\hat{\sigma}^z_j-\bar{c}_{\mu}\hat{a}_{\mu}\hat{\sigma}^z_j+d_{\mu}\hat{a}^{\dagger}_{\mu}\hat{\sigma}^y_j-\bar{d}_{\mu}\hat{a}_{\mu}\hat{\sigma}^y_j,\mathbb{II}\Big]\\
\nonumber &=2 i\sum_jg_{\mu j}^3\Big\{(\hat{a}^{\dagger}_{\mu})^3{\alpha}^{++}_{\mu\mu}\left(t\right)\left(c_{\mu}\hat{\sigma}^y_j-d_{\mu}\hat{\sigma}^z_j\right)+(\hat{a}_{\mu})^3\bar{\alpha}^{++}_{\mu\mu}\left(t\right)\left(\bar{c}_{\mu}\hat{\sigma}^y_j-\bar{d}_{\mu}\hat{\sigma}^z_j\right)\Big\}\\
\nonumber &-i\Big\{\left(\hat{a}_{\mu} (\hat{a}^{\dagger}_{\mu})^2+(\hat{a}^{\dagger}_{\mu})^2\hat{a}_{\mu}\right){\alpha}^{++}_{\mu\mu}\left(t\right)\left(\bar{c}_{\mu}\hat{\sigma}^y_j-\bar{d}_{\mu}\hat{\sigma}^z_j\right)+\left(\hat{a}^{\dagger}_{\mu} (\hat{a}_{\mu})^2+ (\hat{a}_{\mu})^2\hat{a}^{\dagger}_{\mu}\right)\bar{\alpha}^{++}_{\mu\mu}\left(t\right)\left({c}_{\mu}\hat{\sigma}^y_j-{d}_{\mu}\hat{\sigma}^z_j\right)\Big\}\\
 &+2\sum_{j\ne j'}g_{\mu j'}^2g_{\mu j}\Big\{\hat{\sigma}^x_{j'}\hat{\sigma}^z_j\left(\bar{\alpha}^{++}_{\mu\mu}\hat{a}_{\mu}{c}_{\mu}-\alpha^{++}_{\mu\mu}\hat{a}^{\dagger}_{\mu}\bar{c}_{\mu}\right)+\hat{\sigma}^x_{j'}\hat{\sigma}^y_j\left(\bar{\alpha}^{++}_{\mu\mu}\hat{a}_{\mu}{d}_{\mu}-\alpha^{++}_{\mu\mu}\hat{a}^{\dagger}_{\mu}\bar{d}_{\mu} \right)\Big\}\, ,\\
\nonumber &\sum_{ j}g_{\mu j}\Big[c_{\mu}\hat{a}^{\dagger}_{\mu}\hat{\sigma}^z_j-\bar{c}_{\mu}\hat{a}_{\mu}\hat{\sigma}^z_j+d_{\mu}\hat{a}^{\dagger}_{\mu}\hat{\sigma}^y_j-\bar{d}_{\mu}\hat{a}_{\mu}\hat{\sigma}^y_j,\mathbb{III}\Big]\\
\nonumber &=2\sum_{ j,j'}g_{\mu j}^3{B}_{\mathrm{eff};\mu} \left[\left(\hat{\sigma}^y_jc_{\mu}-\hat{\sigma}^z_jd_{\mu}\right)\left(\hat{a}_{\mu} (\hat{a}^{\dagger}_{\mu})^2+ (\hat{a}^{\dagger}_{\mu})^2\hat{a}_{\mu}\right)-\left(\hat{\sigma}^y_j\bar{c}_{\mu}-\hat{\sigma}^z_j\bar{d}_{\mu}\right)\left(\hat{a}^{\dagger}_{\mu} (\hat{a}_{\mu})^2+ (\hat{a}_{\mu})^2\hat{a}^{\dagger}_{\mu}\right)\right]\\
&+2i\sum_{\mu}\sum_{j\ne j'}g_{\mu j'}^2{B}_{\mathrm{eff};\mu} g_{\mu j}\left[\left(c_{\mu}\hat{a}^{\dagger}_{\mu}+\bar{c}_{\mu}\hat{a}_{\mu}\right)\hat{\sigma}^x_{j'}\hat{\sigma}^z_j+\left(d_{\mu}\hat{a}^{\dagger}_{\mu}+\bar{d}_{\mu}\hat{a}_{\mu}\right)\hat{\sigma}^x_{j'}\hat{\sigma}^y_j\right]\, .
\end{align}

Putting all terms together, we find that the single-mode third-order term reads
\begin{align}
\nonumber \hat{\mathcal{A}}_3&=\sum_{j\ne j'} g_{\mu j}\left(\tilde{J}^{xz}_{\mu,j,j'} \hat{\sigma}^x_j\hat{\sigma}^z_{j'}\hat{a}^{\dagger}_{\mu}+\tilde{J}^{xy}_{\mu,j,j'} \hat{\sigma}^x_j\hat{\sigma}^y_{j'}\hat{a}^{\dagger}_{\mu}-\mathrm{H.c.}\right)+\sum_jg_{\mu j}^3 \left[(\hat{a}_{\mu}^{\dagger})^3\left(\alpha^{y\left(3\right)}_{\mu}\hat{\sigma}^y_j+\alpha^{z\left(3\right)}_{\mu}\hat{\sigma}^z_j\right)-\mathrm{H.c.}\right]\\
&+\sum_j g_{\mu j}^3\left[ \left(\hat{a}_{\mu}\left(\hat{a}_{\mu}^{\dagger}\right)^2+\left(\hat{a}_{\mu}^{\dagger}\right)^2\hat{a}_{\mu}\right)\left(\alpha^{y\left(2,1\right)}_{\mu}\hat{\sigma}^y_j+\alpha^{z\left(2,1\right)}_{\mu} \hat{\sigma}^z_j\right)-\mathrm{H.c.}\right]
\end{align}
where
\begin{align}
\tilde{J}^{xz}_{\mu,j,j'}\left(t\right)&=2\left(2\tilde{J}^{zz}_{j,j'}d_{\mu}-\tilde{J}_{j,j'}^{zy} c_{\mu}\right)-2g_{\mu j'}^2\alpha^{++}_{\mu\mu}\bar{c}_{\mu}+2ig_{\mu j'}^2B_{\mathrm{eff};\mu}c_{\mu}\, ,\\
\tilde{J}^{xy}_{\mu,j,j'}\left(t\right)&=-2\left(2\tilde{J}^{yy}_{j,j'}c_{\mu}-\tilde{J}_{j,j'}^{zy} d_{\mu}\right)-2g_{\mu j'}^2\alpha^{++}_{\mu\mu}\bar{d}_{\mu}+2ig_{\mu j'}^2B_{\mathrm{eff};\mu}d_{\mu}\, ,\\
\alpha^{y\left(3\right)}_{\mu}&=2i\alpha^{++}_{\mu\mu} c_{\mu}\, ,\\
\alpha^{z\left(3\right)}_{\mu}&=-2i\alpha^{++}_{\mu\mu} d_{\mu}\, ,\\
\alpha^{y\left(2,1\right)}_{\mu}&=-i\alpha^{++}_{\mu\mu} \bar{c}_{\mu}+2 B_{\mathrm{eff};\mu}c_{\mu}\, ,\\
\alpha^{z\left(2,1\right)}_{\mu}&=i\alpha^{++}_{\mu\mu} \bar{d}_{\mu}-2 B_{\mathrm{eff};\mu}d_{\mu}\, .
\end{align}
Now, in order to find the coefficient of the form $q$ in $\hat{\mathcal{A}}_3$, we replace
\begin{align}
q\to& \frac{1}{2}\int_0^t\Big[q/.\{c\to \frac{i}{2} \cos( B t)e^{-i\delta_{\mu} t},d\to -\frac{i}{2} \sin( B t)e^{-i\delta_{\mu} t},J^{\mu \nu}\to \tilde{J}^{\mu \nu},\alpha^{++}\to \alpha^{++},B_{\mathrm{eff}}\to B_{\mathrm{eff}}\}\\
&+\frac{1}{6}q/.\{c\to \alpha^z,d\to \alpha^y,\tilde{J}^{\mu \nu}\to \check{J}^{\mu \nu},{\alpha}^{++}\to \check{\alpha}^{++},B_{\mathrm{eff}}\to \check{B}_{\mathrm{eff}}\}\Big]\, ,
\end{align}
where the notation $E/.\{\mbox{}\}$ means to apply the replacement rules in braces to the expression $E$.  Using this, the coefficients of the $(\hat{a}^{\dagger}_{\mu})^3$ terms are
\begin{align}
\alpha^{y\left(3\right)}_{\mu}&=\frac{1}{2} 2i\int_0^t dt'\left(\frac{i}{2} \alpha^{++}_{\mu\mu} \left(t'\right)\cos\left( B t'\right)e^{-i\delta_{\mu} t}-\frac{1}{6}\beta_{\mu}\left(t'\right)e^{-i\delta t'}\alpha^z_{\mu}\left(t'\right)\right)\, ,\\
\alpha^{z\left(3\right)}_{\mu}&=-\frac{1}{2} 2i\int_0^t dt'\left(-\frac{i}{2} \alpha^{++}_{\mu\mu} \left(t'\right)\sin\left( B t'\right)e^{-i\delta_{\mu} t}-\frac{1}{6}\beta_{\mu}\left(t'\right)e^{-i\delta t'}\alpha^y_{\mu}\left(t'\right)\right)\, .
\end{align}
These terms are bounded and have the same decoupling points as the first and second-order terms.  The expressions for the terms proportional to $(\hat{a}_{\mu}(\hat{a}_{\mu}^{\dagger})^2+(\hat{a}_{\mu}^{\dagger})^2\hat{a}_{\mu})$ are
\begin{align}
\nonumber \alpha^{y\left(2,1\right)}_{\mu}&=-\frac{i}{2}\int_0^{t} dt' \left(\alpha^{++}_{\mu\mu}\left(t'\right)\left(-\frac{i}{2}\right) \cos\left(B t'\right)e^{i\delta_{\mu} t'}-\frac{1}{6}\beta_{\mu}\left(t'\right)e^{-i\delta t'}\bar{\alpha}^z_{\mu} \left(t'\right)\right)\\
&+\int_0^{t} dt' \left(B_{\mathrm{eff};\mu}\left(t'\right)\frac{i}{2}\cos\left(B t'\right)e^{-i\delta_{\mu} t'}-\frac{i}{12}\left(\beta_{\mu}\left(t\right)e^{i\delta_{\mu} t}-\bar{\beta}_{\mu}\left(t\right)e^{-i\delta_{\mu'} t}\right)\alpha^z_{\mu}\left(t'\right)\right)\\
\nonumber \alpha^{z\left(2,1\right)}_{\mu}&=\frac{i}{2}\int_0^{t} dt' \left(\alpha^{++}_{\mu\mu}\left(t'\right)\left(\frac{i}{2}\right) \sin\left(B t'\right)e^{i\delta_{\mu} t'}-\frac{1}{6}\beta_{\mu}\left(t'\right)e^{-i\delta t'}\bar{\alpha}^y_{\mu} \left(t'\right)\right)\\
&-\int_0^{t} dt' \left(B_{\mathrm{eff};\mu}\left(t'\right)(-\frac{i}{2})\sin\left(B t'\right)e^{-i\delta_{\mu} t'}-\frac{i}{12}\left(\beta_{\mu}\left(t\right)e^{i\delta_{\mu} t}-\bar{\beta}_{\mu}\left(t\right)e^{-i\delta_{\mu'} t}\right)\alpha^y_{\mu}\left(t'\right)\right)
\end{align}
The secular terms from these expressions read
\begin{align}
\alpha^{y\left(2,1\right)}_{\mu}&\approx -\frac{B t}{8\left(\delta_{\mu}^2-B^2\right)^2}\left[\delta_{\mu}+e^{-i\delta_{\mu} t}\left(\delta_{\mu} \cos\left( B t\right)+i B\sin\left(B t\right)\right)\right]\, ,\\
\alpha^{z\left(2,1\right)}_{\mu}&\approx \frac{iB t}{8\left(\delta_{\mu}^2-B^2\right)^2}\left[B+e^{-i\delta_{\mu} t}\left(B \cos\left( B t\right)+i\delta_{\mu} \sin\left(B t\right)\right)\right]\, .
\end{align}

\section{Agreement of the perturbative and non-perturbative Magnus series at lowest order}

In this appendix we check that the perturbative and non-perturbative expansions agree to second order in $g$ but first order in $B$.  That is to say, if we ignore all terms of higher order than $\mathcal{O}\left(B\right)$ in the second-order non-perturbative result, we should recover the second-order perturbative result.  The second-order non-perturbative Magnus propagator is
\begin{align}
e^{i\frac{B}{2}t\sum_j \hat{\sigma}^x_j}e^{\hat{\mathcal{A}}_1+\hat{\mathcal{A}}_2}&\approx e^{i\frac{B}{2}t\sum_j \hat{\sigma}^x_j+\hat{\mathcal{A}}_1+\hat{\mathcal{A}}_2+\frac{1}{2}\left[i\frac{B}{2}t\sum_j \hat{\sigma}^x_j,\hat{\mathcal{A}}_1\right]}\, .
\end{align}
The $\approx$ is an equality up to the order we require, as follows from the Baker-Campbell-Hausdorff formula.  The commutator appearing in the exponential is
\begin{align}
\frac{1}{2}\left[i\frac{B}{2}t\sum_j \hat{\sigma}^x_j,\hat{\mathcal{A}}_1\right]&=\frac{B}{2} t\sum_{\mu j}g_{\mu j}\left[\alpha^z_{\mu}\left(t\right)\hat{a}^{\dagger}_{\mu}-\mathrm{H.c.}\right] \hat{\sigma}^y_j+\mathcal{O}\left(B^2\right)\, ,
\end{align}
and, noting that $B_{\mathrm{eff};\mu}$, $\alpha^{+\pm}_{\mu,\mu'}$, $\tilde{J}^{z,y}_{j,j'}$, and $J^{y,y}_{j,j'}$ are all at least $\mathcal{O}\left(B g^2\right)$ and so beyond the second-order perturbative result, and that $\tilde{J}^{z,z}_{j,j'}=\tilde{J}_{j,j'}(t)+\mathcal{O}\left(B^2\right)$, we find that
\begin{align}
e^{i\frac{B}{2}t\sum_j \hat{\sigma}^x_j}e^{\hat{\mathcal{A}}_1+\hat{\mathcal{A}}_2}&\approx \exp\left(i\frac{B}{2}t\sum_j \hat{\sigma}^x_j-i\sum_{j,j'}\tilde{J}_{j,j'}(t)\hat{\sigma}^z_j\hat{\sigma}^z_{j'}+\sum_{\mu j}g_{\mu j} \left[\alpha_{\mu}(t)\hat{a}^{\dagger}_{\mu} -\mathrm{H.c.}\right]\hat{\sigma}^z_j+\sum_{\mu j}g_{\mu j} \left[\gamma_{\mu}(t)\hat{a}^{\dagger}_{\mu} -\mathrm{H.c.}\right]\hat{\sigma}^y_j\right)\, ,
\end{align}
where we have used the fact that $\lim_{B\to 0}\alpha^z_{\mu}(t)=\alpha_{\mu}(t)$.  The spin-boson coupling $\gamma_{\mu}(t)$ is the first-order coefficient of $\alpha^y_{\mu}(t)$ plus the factor of $B t\alpha^z_{\mu}\left(t\right)$ from the BCH commutator above.  To wit, we find
\begin{align}
\gamma_{\mu}(t)&=\frac{B}{2\delta_{\mu}^2}\left(i+e^{-i\delta_{\mu} t}\left(\delta_{\mu} t-i\right)\right)+\frac{Bt}{4\delta_{\mu}}\left(1-e^{-i\delta_{\mu} t}\right)=\frac{B}{4\delta_{\mu}^2}\left[\delta_{\mu} t\left(1-e^{-i\delta_{\mu} t}\right)+2i\left(1-e^{-i\delta_{\mu} t}\right)\right]\, ,
\end{align}
as was found for the perturbative result.  Hence, the two expansions agree to the specified order.

\section{Time-ordered dynamics of the effective Hamiltonian Eq.~\eqref{eq:CanonEff} resulting from the canonical transformation approach}
\label{app:Can}

Working through the first two orders of the Magnus expansion of Eq.~\eqref{eq:CanonEff} without any of the terms in the correction series, we find
\begin{align}
\hat{\mathcal{U}}(t)&\approx \exp\left(-i\sum_{j\ne j'}\tilde{J}_{j,j'}\left(t\right)\hat{\sigma}^z_j\hat{\sigma}^z_{j'}-i \hat{H}_B t+\hat{\mathcal{A}}^{(c)}_2\right)\, ,
\end{align}
where the second-order term is
\begin{align}
\nonumber \hat{\mathcal{A}}^{(c)}_2(t)&=-\sum_{j\ne j'}\frac{1}{2}\int_0^tdt' \left[-i\tilde{J}_{j,j'}\left(t'\right)\hat{\sigma}^z_j\hat{\sigma}^z_{j'},-i\frac{B}{2}\sum_{k} \hat{\sigma}^x_k\right]=i \frac{B}{2}\sum_{j\ne j'} \int_0^{t} dt' \tilde{J}_{j,j'}\left(t'\right)\hat{\sigma}^y_j\hat{\sigma}^z_{j'}\, .
\end{align}
Hence, even in the absence of boson effects, the effective Hamiltonian $\hat{H}_{\mathrm{eff}}$ does not strictly correspond to a TFIM with spin-spin couplings given by $\tilde{J}_{j,j'}(t)$.

Let us now look at the first two terms in the ``correction series" given by the commutators in Eq.~\eqref{eq:CanonEff}.  The first-order correction $\hat{C}_1(t)=-\left[\hat{\mathcal{A}}^{(p)}_1(t),\hat{H}_B\right]$ is
\begin{align}
\nonumber &\hat{C}_1(t)=\sum_{\mu} \left[\sum_{j}g_{\mu j}\left(\alpha_{\mu}(t)\hat{a}^{\dagger}_{\mu}-\mathrm{H.c.}\right)\hat{\sigma}^z_{j},B\sum_k\hat{\sigma}^x_j\right]=-i\sum_{\mu}\sum_{j}\frac{B}{2}g_{\mu j}\left(\alpha_{\mu}(t)\hat{a}^{\dagger}_{\mu}-\mathrm{H.c.}\right)\hat{\sigma}^y_j\, ,
\end{align}
which gives a spin-dependent force along the $\hat{\sigma}^y$ direction.  The next-order term is
\begin{align}
\hat{C}_2(t)&=-\frac{1}{2}\Big[\sum_{\mu'}\sum_{j'}g_{\mu' j'}\left(\alpha_{\mu'}(t)\hat{a}^{\dagger}_{\mu'}-\mathrm{H.c.}\right)\hat{\sigma}^z_{j'},\sum_{\mu}\sum_{j}\frac{B}{2}g_{\mu j}\left(\alpha_{\mu}(t)\hat{a}^{\dagger}_{\mu}-\mathrm{H.c.}\right)\hat{\sigma}^y_j\Big]\, ,\\
\nonumber &=i\sum_{\mu,\mu'}\sum_j\frac{Bg_{\mu j}g_{\mu' j}}{4}\hat{\sigma}^x_j\left(\alpha_{\mu'}(t)\hat{a}^{\dagger}_{\mu'}-\mathrm{H.c.}\right)\left(\alpha_{\mu}(t)\hat{a}^{\dagger}_{\mu}-\mathrm{H.c.}\right)\,.
\end{align}
Here, we note that the second order term in the Magnus series generated from $\hat{C}_1(t)$, i.e., $-\frac{1}{2}\int_0^t dt' \int_0^{t'} dt''\left[\hat{C}_1(t'),\hat{C}_1(t'')\right]$ gives spin-spin interactions along the $yy$ direction, and the first order term in the Magnus expansion of $\hat{C}_2(t)$, i.e. $-i\int_0^tdt' \hat{C}_2(t')$, gives rise to the thermally dependent effective magnetic field, denoted $B_{\mathrm{eff}}$ in the frame of the non-perturbative calculation.  Hence, while the correction series appears to only involve spin-boson couplings, the time-ordering of its terms can also produce effective spin-spin interactions at the same order in $g/\delta$ as the Ising spin-spin interactions.

\end{document}